\documentclass[manuscript,nonacm]{acmart}

\usepackage[T1]{fontenc}
\usepackage[utf8]{inputenc}
\usepackage{setspace}
\usepackage{lineno}
\usepackage{graphicx}
\usepackage{booktabs}
\usepackage{threeparttable}
\usepackage{longtable}
\usepackage{tabularx}
\usepackage{array}
\usepackage{amsmath}
\usepackage{enumitem}
\usepackage{caption}
\usepackage{placeins}
\usepackage{hyperref}

\usepackage{multirow}
\usepackage{pdflscape}
\usepackage{xcolor}
\usepackage{soul}
\usepackage{tcolorbox}
\usepackage[normalem]{ulem}

\newcolumntype{L}[1]{>{\raggedright\arraybackslash}p{#1}}
\newcolumntype{C}[1]{>{\centering\arraybackslash}p{#1}}
\newcolumntype{Y}{>{\raggedright\arraybackslash}X}

\sethlcolor{yellow} %

\AtBeginDocument{%
  }

\setcopyright{acmlicensed}
\copyrightyear{2018}
\acmYear{2018}
\acmDOI{XXXXXXX.XXXXXXX}
\acmConference[Conference acronym 'XX]{Make sure to enter the correct
  conference title from your rights confirmation email}{June 03--05,
  2018}{Woodstock, NY}
\acmISBN{978-1-4503-XXXX-X/2018/06}

\begin{document}

\title{Metacognitive Scaffolds Create Productive Friction for AI-Assisted Knowledge Work}

\author{Viktor Kewenig}
\authornote{Both authors contributed equally to this research.}
\affiliation{%
  \institution{Microsoft Research}
  \city{Cambridge}
  \country{United Kingdom}
}
\email{t-vkewenig@microsoft.com}

\author{Lev Tankelevitch}
\authornotemark[1]
\affiliation{%
  \institution{Microsoft Research}
  \city{Cambridge}
  \country{United Kingdom}
}
\email{lev.tankelevitch@microsoft.com}

\author{Sean Rintel}
\affiliation{%
  \institution{Microsoft Research}
  \city{Cambridge}
  \country{United Kingdom}}
\email{serintel@microsoft.com}

\author{Advait Sarkar}
\affiliation{%
  \institution{Microsoft Research}
  \city{Cambridge}
  \country{United Kingdom}
  }
\email{advait@microsoft.com}

\author{Abigail Sellen}
\affiliation{%
  \institution{Microsoft Research}
  \city{Cambridge}
  \country{United Kingdom}}
\email{asellen@microsoft.com}

\begin{abstract} %

Generative AI imposes new metacognitive demands on knowledge workers as their work shifts from production to planning, evaluation, and orchestration. In a randomized, between-subjects experiment on AI-assisted writing (297 workers, 594 essays), we test dynamically-generated metacognitive scaffolds supporting goal specification and task decomposition. We find that scaffolds improve writing quality, with this boost persisting in short-lag transfer when scaffolds are removed. We also find evidence of improved topic recall three days later. These effects are consistent with observed changes in how people learn to prompt AI systems after exposure to scaffolding. However, scaffolded participants rate the AI as less helpful and more cognitively demanding even as they perform better, suggesting that subjective ease can understate the value of support. Our findings make the case for metacognitive scaffolds in AI-assisted knowledge work, revealing a productive friction worth designing for as workers increasingly delegate work to agentic AI.

\end{abstract}

\renewcommand{\shortauthors}{Kewenig and Tankelevitch et al.}

\begin{teaserfigure}
 \includegraphics[width=\textwidth]{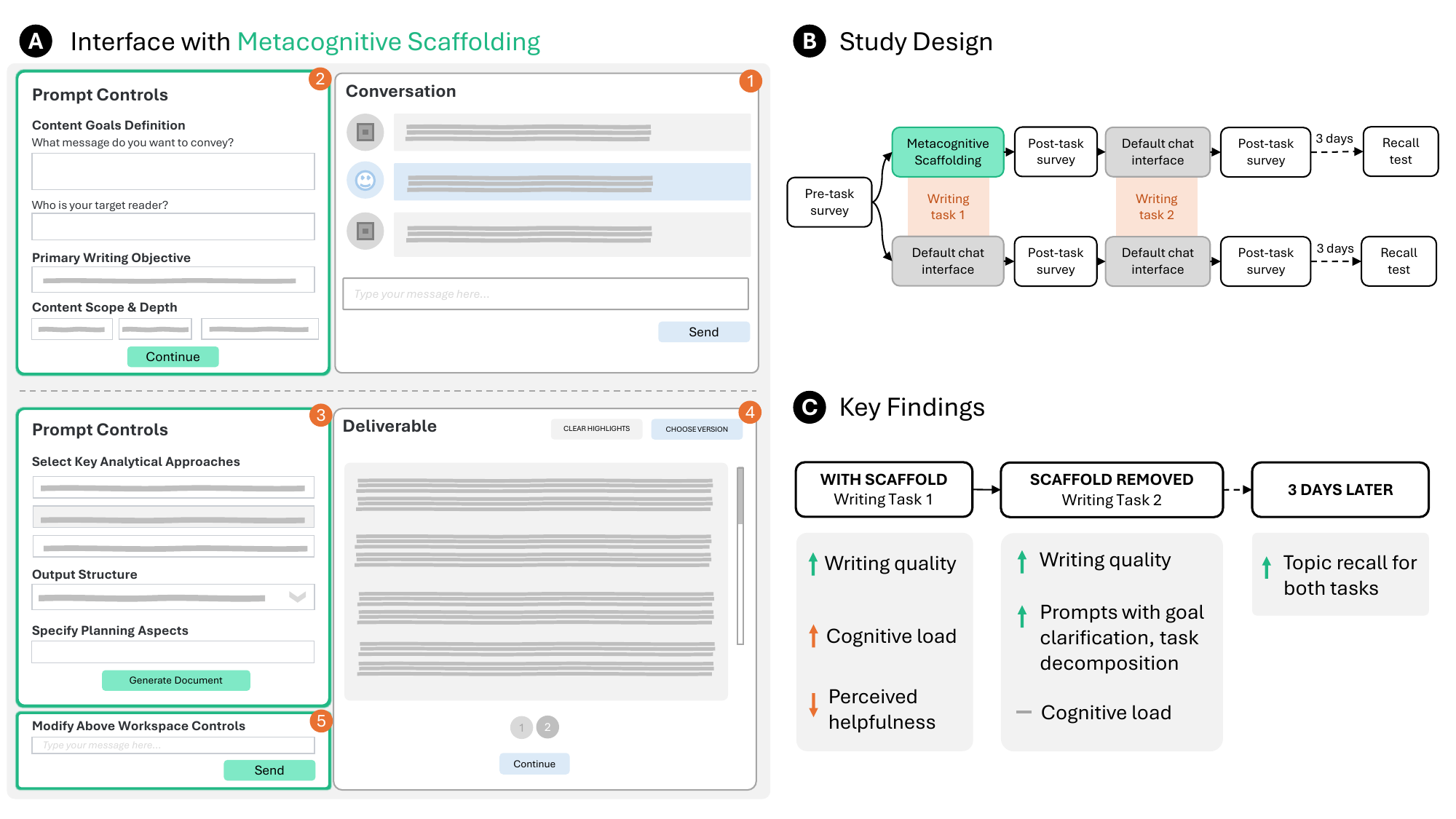}
  \caption{\textbf{A}, The AI-assisted writing interface with the metacognitive scaffold: after an initial request (1), dynamically generated controls prompt goal specification (2) and task decomposition (3) before a draft is generated (4); controls could be revised in natural language (5).\textbf{B}, Randomized experimental design: 297 knowledge workers wrote two briefs, with the scaffold available only for one group in Task~1, followed by a recall test three days later. \textbf{C}, Key findings: better briefs with the scaffold, at the cost of higher load and lower perceived helpfulness; the quality advantage and more deliberate prompting persisted after removal; evidence of better topic recall three days later.}
  \Description{Description.}
  \label{fig:teaser}
\end{teaserfigure}

\maketitle

\section{Introduction}\label{sec:introduction}

Generative AI (hereafter, AI) is rapidly reshaping knowledge workflows across professional, educational, and creative settings. Increasing experimental evidence demonstrates that access to large language models (LLMs) can raise both productivity and output quality for tasks such as writing, analysis, and problem-solving~\cite{Brynjolfsson2025GenAIWork,Noy2023,DellAcqua2023JaggedFrontier,Dillon2025EarlyImpactsCopilot,Dillon2025WorkPatterns}. Yet these gains come with a fundamental shift in the nature of human contribution: rather than executing work directly, the human role increasingly centers on steering and orchestrating model behavior, including specifying goals, decomposing tasks, and evaluating generated output~\cite{sarkar_exploring_2023,Tankelevitch2024,palani_evolving_2024,he_rebalancing_2023,Li2024,nguyen2024collaboration,Lo2023,Kasneci2023}. This redistribution of effort imposes \textit{metacognitive} demands on people that are poorly supported by current AI interfaces~\cite{Tankelevitch2024,Subramonyam2024GulfEnvisioning}. Accordingly, many workers struggle with adopting AI tools or use them ineffectively, slowing their workflows or hampering output quality~\cite{Lee2025,DellAcqua2023JaggedFrontier}. Metacognitive demands intensify further as AI systems become more agentic, with autonomous systems executing multi-step plans, invoking tools, and iterating towards goals on workers' behalf~\cite{sarkar-s_aiagents_2026, passi_agentic_2025, dhanorkar_human_2026}. 

Alongside these challenges of steering and orchestration, using AI increasingly entails offloading material production to the model itself. Workers may no longer write sentences, generate code, or produce other artifacts directly, a shift demonstrated in both experimental and observational studies~\cite{shaw_thinkingfast_2026,Lee2025,kim_i_2026}. 
If workers are increasingly specifying goals or task breakdowns for AI, then these metacognitive processes by definition become a key part of how workers engage with and understand their own work. Consequently, insufficient support for these processes risks not only poorer steering of AI systems, but also reduced cognitive engagement with the work itself. 

Recent research has begun to address this by exploring AI interfaces that scaffold the articulation of evolving goals~\cite{kim_intentflow_2026, Gmeiner2025MetacognitiveAgents, drosos_dynamic_2025}, tracking goals across conversations~\cite{Coscia2025OnGoal}, decomposing tasks into sub-tasks~\cite{yen_coladder_2024,feng_cocoa_2026,kazemitabaar_improving_2024, drosos_dynamic_2025}, and other metacognitive processes important for human-AI interaction. Yet despite this progress, prior work has primarily examined proximal interaction outcomes, such as usability and model steerability. Value in knowledge work depends on more distal outcomes, including output quality, whether benefits persist in the short or long term once support is removed, and cognitive engagement with the work itself. We make progress on all three.

We present a randomized between-subjects experiment in AI-assisted writing (297 participants, 594 essays), evaluating the impact of metacognitive scaffolding on multiple outcomes key to AI-assisted knowledge work: output quality, prompting ability, cognitive load, and topic recall of work material, as one key aspect of cognitive engagement. We focus on \textit{goal specification} and \textit{task decomposition} as the primary targets for metacognitive scaffolding. %
These processes sit at the beginning of AI-assisted workflows and are therefore key for human-GenAI interaction, as the quality of goal specification and task structure determines how effectively all downstream activities can proceed~\cite{Tankelevitch2024,tankelevitch2026goals}. As such, we implemented metacognitive scaffolds as AI-generated, context-specific interface controls for goal specification and task decomposition (the interface and interaction sequence are described in \autoref{sec:research-probe}).

We first compare the scaffolds' impact on written output quality, cognitive load, and other aspects of user experience to a baseline interface without scaffolds. Unlike prior work, we then test whether the scaffolds' impact on human-AI interaction persists into short-lag transfer when support is removed. That is, in a subsequent writing task in the same session where we remove the scaffolds, we evaluate how scaffold exposure affects written output quality and prompting behavior. Finally, to test one aspect of how scaffolds affect cognitive engagement with task material, we follow up with participants three days later to assess their topic recall for both writing tasks. 

We find that metacognitive scaffolding improves AI-assisted output quality, helping participants produce written outputs that are more actionable, better organized, and more clearly framed for the task. This effect persists after removal of support. We also find that participants learn to prompt the AI toward clearer goals and stronger task decomposition, offering a potential explanation for this carryover effect. Finally, we find evidence for improved topic recall. However, all of these gains carry a subjective cost: despite better performance, scaffolded participants rate the AI as less helpful and report an increase in immediate post-task cognitive load. We argue that this \textit{productive friction}---support that increases felt effort and dampens perceived helpfulness while improving outcomes---is a crucial challenge in designing AI systems for knowledge work.

In sum, we make the following contributions with respect to dynamic metacognitive scaffolding for goal specification and task decomposition in AI-assisted writing:

\begin{enumerate}
    \item We show that dynamic metacognitive scaffolding improves writing quality and prompt quality, and is associated with improved topic recall, and that these improvements survive short-lag transfer when scaffolds are removed.

    \item We find that the benefits of dynamic metacognitive scaffolding come with key trade-offs against perceived cognitive load and helpfulness.
    
    \item We offer research and design implications for advancing the development of metacognitive support for AI-assisted knowledge work.
    
\end{enumerate}

\section{Related Work}\label{sec:relatedwork}

\subsection{Generative AI's Reshaping of Knowledge Workflows}\label{subsec:reshaping}

Generative AI has rapidly diffused through knowledge work~\cite{Bick2026RapidAdoption}. Across a representative sample of ChatGPT conversations, practical guidance, information seeking and writing account for nearly 80\% of use, and writing dominates work-related messages~\cite{Chatterji2025HowPeopleUseChatGPT}; similar patterns occur across millions of enterprise Microsoft Copilot Chat ~\cite{Counts2026Enterprise} and Claude conversations~\cite{Handa2025EconomicTasks}. This applicability is broad but shallow, concentrated in creating, processing and communicating information rather than end-to-end automation~\cite{Tomlinson2025WorkingWithAI}.

Controlled experiments across various professions find that LLM access cut writing time and raised quality among professionals~\cite{Noy2023}, accelerated customer support resolutions~\cite{Brynjolfsson2025GenAIWork}, and increased completed programming tasks ~\cite{Cui2026ThreeFieldExperiments}. The gains are uneven, falling away beyond a ``jagged frontier'' of suitable tasks~\cite{DellAcqua2023JaggedFrontier}, and their distribution across expertise depends on the tool, with different studies finding largest gains among less experienced workers ~\cite{Brynjolfsson2025GenAIWork,Cui2026ThreeFieldExperiments,Cruces2026EducationGaps}, more experienced workers ~\cite{Daniotti2026WhoIsUsingAI}, and contradictory effects between perception and reality~\cite{Becker2025Slowdown}.

In some cases the composition of work changes, and in others simply the volume of work, depending on what the tool does. A two-year study of inline code completion found developers shifting activity toward core coding and away from non-core project management, interpreted as a move from collaborative toward
independent work~\cite{Hoffmann2025NatureOfWork}. That reallocation runs between joint and solo work rather than between producing and directing. Nor does reallocation follow automatically from access: a large-scale field experiment among knowledge workers found time savings \textit{without} detectable task recomposition~\cite{Dillon2025WorkPatterns}.

Where the model produces the entire artifact rather than completing small portions, however, the axis itself changes. Conceptually, the shift has been described as one from material production to \textit{critical integration}~\cite{sarkar_exploring_2023}. Where automation can be applied within a workflow influences which forms of human contribution are considered most valuable. In this framing, people contribute primarily by specifying objectives, planning work, selecting and combining outputs, and critically evaluating what the model provides. This shift raises obvious questions about people's cognitive engagement with their work: what is offloaded and what cognitive tasks remain essential, or become \textit{more} important, for AI-assisted work? We discuss each in turn.   

\subsection{Cognitive Offloading and Engagement}\label{subsec:offloading}

Delegating production to AI can entail \textit{cognitive offloading}: recruiting external resources to reduce a task's cognitive demands~\cite{RiskoGilbert2016}. Prior research has found that among knowledge workers, greater confidence in AI was associated with less perceived effort for critical thinking, and the remaining effort shifted from producing information toward verifying and integrating it~\cite{Lee2025}. Workers may not notice that such offloading is happening. They may misjudge the division of labour: in real world task-focused conversation logs, the model is credited with 11-26\% of `goal-shaping' contributions, revealing an over-estimation in users of their own contributions~\cite{kim_i_2026}. They may also misjudge the result's quality: across studies, participants tend to over-rely on AI assistance, inappropriately adopting its output with little scrutiny, while gaining self-confidence~\cite{shaw_thinkingfast_2026,colombatto_metacognition_2025,Fernandes2026NoneTheWiser}. 

Recent work has begun to measure the effects of offloading. Across recent large field studies of students, AI assistance tended to improve performance when available, e.g., during homework or non-proctored tests, but impaired performance when withdrawn, such as in closed-book exams~\cite{Bastani2025,Stromberg2026LearningPenalty,Rismanchian2026FasterCompletion}. Similar effects appear for memory of authorship~\cite{Zindulka2026MemoryGap}, for metacognitive
accuracy~\cite{Fernandes2026NoneTheWiser,Marcoccia2026SuspendJudgment}, and for professional skill such as clinical diagnostic detection~\cite{Budzyn2025Deskilling}. Such erosion can be asymptomatic: in one high-stakes workplace, early operational gains masked a dulling of expert judgment that only later surfaced as skill atrophy~\cite{Ehsan2026IntuitionRust}. On the other hand, using records from a large U.S. university, a study comparing AI-exposed to non-exposed courses found no significant effect of ChatGPT's arrival on grades, and no difference in self-reported understanding~\cite{Dumlao2026Availability}. Caution in interpreting these prior results is warranted, as publication bias and methodological heterogeneity across studies makes any overarching verdict on AI and learning premature~\cite{Bartos2026MetaMeta}. The evidence also comes chiefly from learning settings rather than knowledge work, and seldom measures cognitive engagement and output quality within a single design, a gap we address.

In knowledge work, cognitive engagement is a complex constellation of elements at both the individual level (e.g. critical thinking~\cite{zhi_investigating_2026,Lee2025}) and collaborative level (e.g. articulation work~\cite{riedl_cognitive_2026,strauss1988articulation,schmidt1992articulation}). At the individual level, memory (often measured as topic recall or source attribution) is a key aspect, and also a key fragility. 
Prior research has found that AI assistance may be detrimental to memory~\cite{Bastani2025,Zindulka2026MemoryGap,zhi_investigating_2026,chen_more_2025,kreijkes_effects_2025}. For example,~\citet{Zindulka2026MemoryGap}'s participants generated and elaborated ideas with and without AI. One week later, correct source attribution had fallen significantly, with the largest decline in mixed human-AI workflows. Other work has found negative impacts of AI assistance on content recall and comprehension \cite{zhi_investigating_2026, kreijkes_effects_2025}. Topic recall can function as a simple indicative measure of individual cognitive engagement.

However, cognitive offloading is not monolithic, and nor are its effects on memory and other aspects of cognition. The type of cognitive tasks that people offload and how they direct their effort matters for downstream impact. For example, when given an AI writing tool that provided exposure to high-quality AI-generated examples, participants exerted less effort while practicing, but then wrote better cover letters in later unaided tests, outperforming both web search and personalized feedback from human editors~\cite{Lira2026CoachNotCrutch}. In studies of programmers~\cite{Shen2026SkillFormation} and students~\cite{contractor_experimental_2026}, those who delegated wholesale lost conceptual understanding, yet certain `augmentation' interaction patterns actually preserved learning. Guardrails or engagement techniques that structure learning interactions reversed the harm of unguarded AI assistance~\cite{Bastani2025} and improved learning without withdrawing assistance altogether ~\cite{Kazemitabaar2025CognitiveEngagement}. 

The timing of cognitive engagement also matters. Requiring students to generate ideas before consulting an LLM conferred no immediate advantage but the largest gains on a later unassisted task~\cite{Wong2026ThinkFirst}. In knowledge work tasks with sufficient time available, avoiding unfettered LLM access specifically when \textit{initiating} work preserved critical thinking and memory retention of materials, while improving overall quality~\cite{zhi_investigating_2026}. Overall, scaffolding early-stage thinking during AI-assisted work may be effective in driving cognitive engagement with work material. This early stage entails specifying goals and decomposing tasks for AI, key metacognitive demands that are the counterpart of offloading, and the focus of our study.

\subsection{Metacognitive Demands of Generative AI}\label{subsec:metacognitivedemands}

The effort AI displaces from producing information reappears as activities necessary for effective delegation. For example, workers experienced in AI are distinguished chiefly by planning in their opening messages~\cite{sarkar-s_aiagents_2026} and more robust verification processes~\cite{cynthia_are_2026}. These activities recruit metacognitive processes, including self-monitoring, confidence calibration, and cognitive flexibility~\cite{Tankelevitch2024,fugener_cognitive_2022}. The shift from material production to critical integration therefore imposes metacognitive demands on users as they decide \textit{what} to delegate to AI~\cite{Tankelevitch2024,sarkar_exploring_2023,fugener_cognitive_2022,jussupow_augmenting_2021}, and then grapple with the open-ended loop of prompting--evaluating--interaction, i.e., \textit{how} to delegate to AI~\cite{Subramonyam2024GulfEnvisioning,sarkar_what_2022,liu_what_2023}. Recent work has further demonstrated the role of these metacognitive processes in human--AI interaction~\cite{singh_exploring_2025,Lee2025,tankelevitch_understanding_2025}, including in reasoning~\cite{Fernandes2026NoneTheWiser,bo_rely_2025}, planning~\cite{colombatto_metacognition_2025}, information seeking~\cite{ramesh_metacognitive_2026,mittal_learning_2026}, software engineering~\cite{sarkar-s_aiagents_2026}, and data science~\cite{drosos_its_2024}. These demands may intensify as AI systems take on longer, multi-step workflows. Longer autonomous task horizons leave users with fewer opportunities to intervene, while increasing the consequences of their initial specifications and later verification decisions~\cite{Kwa2025TimeHorizon,passi_agentic_2025,dhanorkar_human_2026}. Addressing the metacognitive demands of AI through interface support therefore remains a priority~\cite{Tankelevitch2024}. 

\subsection{Metacognitive Scaffolds for Goal and Task Specification in AI-Assisted Workflows}\label{subsec:metacognitivescaffolds}

Research on metacognitive scaffolds has begun to explore the range of activities and associated metacognitive processes that span AI-assisted workflows, from goal/intent specification~\cite{suh_luminate_2024,gmeiner_intent_2025,Coscia2025OnGoal,drosos_dynamic_2025,kim_intentflow_2026} and task decomposition~\cite{wu_ai_2022,
wu_promptchainer_2022,feng_cocoa_2026,yen_coladder_2024,Kazemitabaar2024}, to verification and appropriate reliance~\cite{yen_coladder_2024,Kazemitabaar2024, ghosh_experimental_2026,kim_i_2026, bo_rely_2025, KimReliance2025, cheng_relic_2024, ashktorab_emerging_2025, arun_encouraging_2026}. Our study focuses on goal specification and task decomposition as two key interrelated activities that most commonly initiate and structure AI-assisted workflows \cite{Tankelevitch2024,tankelevitch2026goals}.      

\subsubsection{Goal and Intent specification} 

Prior work has explored several ways of helping users articulate, explore, clarify, and manage their goals and intents across multi-turn LLM interactions~\cite{kim_intentflow_2026}. In co-creation workflows, \textit{Luminate}~\cite{suh_luminate_2024} surfaces a multi-dimensional design space from users' prompts, helping them discover relevant combinations, while \textit{Intent Tagging}~\cite{gmeiner_intent_2025} represents intentions as granular, reusable units that support flexible, non-linear expression and steering. \textit{OnGoal}~\cite{Coscia2025OnGoal} surfaces goals throughout a conversation, using an LLM-assisted pipeline to infer, merge, and evaluate goals against model responses. \textit{Dynamic Prompt Middleware}~\cite{drosos_dynamic_2025} augments users' prompts through dynamically generated interface controls, and \textit{IntentFlow}~\cite{kim_intentflow_2026} similarly extracts goals and intents into editable components that let users iteratively articulate, verify, and curate their intents. In line with metacognition research, \cite{Molenaar2011,Molenaar2013}, a key design pattern across this work is dynamic, activity-contingent support that can be tied to the user's current problem state and that can therefore be more effective than generic static scaffolds ~\cite{drosos_dynamic_2024, kim_intentflow_2026}. 

\subsubsection{Task Decomposition} 

A complementary line of work externalizes task structure to help users decompose complex goals, clarify their intentions, and remain aware and in control of their intent at each step. AI Chains~\cite{wu_ai_2022,wu_promptchainer_2022} introduced editable multi-step LLM workflows, an approach echoed in Cocoa's agentic workflows~\cite{feng_cocoa_2026} and CoLadder's links between hierarchical intentions and corresponding code~\cite{yen_coladder_2024}.~\citet{Kazemitabaar2024} compared phasewise decomposition (separating assumptions, planning, and code into larger phases) with stepwise decomposition into smaller subgoals, finding that both facilitated steering and verification. Brickify similarly externalizes design intent as manipulable visual tokens, allowing users to specify both the desired elements and how they should be composed; our scaffold applies a related principle to goals and task structure in writing~\cite{shi_brickify_2025}.

\subsection{Synthesis and Current Study} 

Prior work demonstrates how scaffolds can improve users' goal and intent discovery, clarification, task decomposition, planning, and steering of model behavior. Beyond these UX and interaction outcomes, a few studies suggest that scaffolds can increase output quality in coding~\cite{yen_coladder_2024} and other tasks~\cite{wu_ai_2022}, and even improve code comprehension~\cite{yen_coladder_2024}. They also reveal a tension around cognitive load, with mixed findings about whether scaffolds' added options and interaction complexity increase or decrease load compared to common chat interfaces. Together, this work establishes both the promise of metacognitive support for human-AI interaction and its central design challenges. 

Our work builds on these design approaches and their formative insights, drawing on the positive results of dynamically-generated interface controls~\cite{drosos_dynamic_2025} to support goal specification and task decomposition. We ask how such scaffolds impact the value of outcomes in knowledge work. As noted above, prior evaluations have largely measured proximal interaction outcomes. Value, however, depends on more distal work outcomes that remain largely untested: objective output quality, whether scaffolds shape users' interaction patterns in subsequent tasks, and, as one key aspect of cognitive engagement, users' topic recall of work material. We take a step toward these outcomes, while acknowledging that durable skill change and in-depth cognitive engagement entail complexities well beyond what can be accomplished by an individual worker in a single session. Methodologically, we complement the largely small-scale, within-subjects prior work with a large-scale, between-subjects evaluation designed to isolate scaffolds' causal effects with greater generalizability.

We examine the impact of metacognitive scaffolding for goal specification and task decomposition through the following research questions. Compared to a standard chat-and-canvas interface:

\begin{itemize}
    \item \textbf{RQ1:} Do scaffolds improve AI-assisted writing quality, and do effects carry over when working with AI in the absence of scaffolds?
    \item \textbf{RQ2:} As a mechanism for potential carry-over effects, do scaffolds change how people prompt AI systems in the absence of scaffolds?  
    \item \textbf{RQ3:} Do scaffolds improve people's topic recall of work materials, as a component of cognitive engagement?
    \item \textbf{RQ4:} How do scaffolds impact perceived helpfulness, sense of control, and cognitive load?
\end{itemize}

\section{Methods}\label{sec:methods}

\begin{figure}
    \centering
\includegraphics[width=0.97\linewidth]{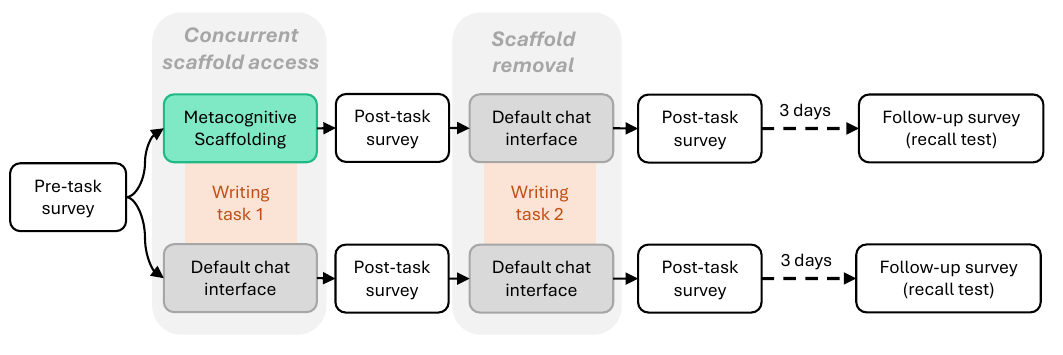}
\caption{Participant flow in the study.}
\label{fig:flow}
\end{figure}

We conducted a randomized between-subjects experiment focusing on AI-assisted writing (\autoref{fig:flow}). In a main study session, participants completed two comparable professional-writing tasks in counterbalanced order. In Task~1, participants assigned to the \textit{metacognitive scaffold} condition used an interface with dynamically generated controls to support goal specification and task decomposition, whereas \textit{control} group participants used the otherwise identical chat-and-canvas interface without these panels. This contrast therefore evaluates the effect of concurrent scaffold access. In Task~2, participants in both conditions used the unscaffolded interface. This contrast tests short-lag transfer after scaffold removal. 

Participants completed a baseline \textit{pre-task survey} measuring AI usage, metacognition and writing skills, and a \textit{post-task survey} after each task: after Task~1, metacognitive self-report, perceived helpfulness, control and cognitive load; after Task~2, cognitive load only. Approximately three days later, they completed a \textit{follow-up survey} assessing topic recall for both writing tasks. We also collected and analyzed their written outputs and prompts. The study was not preregistered.

\subsection{Participants and Recruitment}

Following ethics authorization\footnote{Anonymized for review}, we recruited participants from Prolific. Participants provided informed consent electronically before taking part. A total of 342 participants entered the study and were compensated £20/hour. Of these, 44 participants did not submit a final essay or their essay could not be recovered; one participant's baseline questionnaire was not saved due to technical difficulties. The final analytic sample for the primary analyses of writing quality and prompting therefore comprised 297 participants who completed both writing tasks (control = 144; treatment = 153) and 594 essays. Survey analyses used all participants with complete data on the relevant baseline and post-task items (279--295 per item; \autoref{tab:metacognition-items}). Cognitive-load analyses used 277 participants with measurements at both analyzed timepoints. Despite a reward being announced at the end of the session (a chance to win one of 20 extra Prolific rewards of over \pounds30 for completing the follow-up), attrition for the follow-up survey was high: 159 participants returned (we address this in \autoref{subsubsec:recall-missingness}). All participants were based in the United States and reported English as their first language (\autoref{tab:demographics} summarizes participant characteristics).

\renewcommand{\arraystretch}{1.3}
\begin{table}[htbp]
\centering
\footnotesize
\caption{Participant characteristics of final analytic sample. Values are
$n$ (\%) unless otherwise indicated. IQR, interquartile range; SD, standard
deviation.}
\label{tab:demographics}

\begin{tabularx}{\linewidth}{@{}L{0.25\linewidth}Xrr@{}}
\toprule
& & Control ($n=144$) & Treatment ($n=153$) \\
\midrule
Age
& Mean (SD)                    & 34.2 (12.0) & 33.6 (11.9) \\
& Median [IQR]                 & 31 [26, 41] & 31 [24, 42] \\ 
\midrule
Gender
& Male                         & 71 (49.3\%) & 78 (51.0\%) \\
& Female                       & 71 (49.3\%) & 69 (45.1\%) \\
& Non-binary or other          & 2 (1.4\%)   & 6 (3.9\%) \\
\midrule
Country
& USA                          & 144 (100\%) & 153 (100\%) \\
\midrule
First language
& English                      & 144 (100\%) & 153 (100\%) \\
\midrule
Highest educational qualification
& High school                  & 18 (12.5\%) & 30 (19.6\%) \\
& Bachelor's degree            & 74 (51.4\%) & 88 (57.5\%) \\
& Master's degree              & 29 (20.1\%) & 17 (11.1\%) \\
& Doctoral or professional degree
                               & 23 (16.0\%) & 18 (11.8\%) \\
\midrule
AI usage frequency
& Monthly & 10 (6.9\%) & 9 (5.9\%) \\
& Weekly  & 58 (40.3\%) & 85 (55.6\%) \\
& Daily   & 76 (52.8\%) & 59 (38.6\%) \\
& Missing & 0 (0.0\%) & 0 (0.0\%) \\
\bottomrule
\end{tabularx}
\end{table}
\renewcommand{\arraystretch}{1}

\subsection{Research Probe and Interface Design}\label{sec:research-probe}

In both conditions, participants completed each task in a browser-based writing environment consisting of a chat panel connected to GPT-5.3 with medium reasoning effort and an editable canvas panel. Participants could prompt the AI freely, insert generated text into the document, and edit manually at any point; there was no minimum and no limit on the number of turns. AI-generated changes were indicated via text highlighting, and participants could toggle between different version histories of their document. After completing the main AI-assisted writing phase, participants were able to do a final manual review of the document before submitting. 

\subsubsection{Dynamic Metacognitive Scaffold}

The scaffold was motivated by work on dynamic prompt middleware, showing that context-sensitive interface controls can increase steering precision, perceived control \textit{and task reflection} when interacting with AI~\cite{drosos_dynamic_2025}. Broadly, its metacognitive support focused on intent articulation and exploration as defined by~\citet{kim_intentflow_2026}. Concretely, it was implemented as a two-phase \textit{dynamic} interface (\autoref{app:interface}).

After the participant entered a high-level request, the system rendered a \textit{goal specification} panel that externalized the rhetorical problem through editable fields for the message to convey, target reader, desired reader outcome, primary writing objective and desired scope/depth (see figure \ref{fig:teaser}, panels A.1 and A.2). This phase targeted explicit problem representation, reducing ambiguity about what would count as a good response. Once those inputs were specified, the interface rendered a \textit{task decomposition} panel that generated task-specific controls for analytical approaches, writing structure and critical considerations to address in the brief. This phase targeted actionability and verification (see figure \ref{fig:teaser}, panels A.3 and A.4). Participants could also request modifications to these controls in natural language through a dedicated workspace control panel (see figure \ref{fig:teaser}, panel A.5). This mapping from clarified goals to decomposed sub-goals follows classic theories of hierarchical problem solving~\cite{NewellSimon1972,Shepherd2001} and recent human--AI interface work on decomposable steering~\cite{Kazemitabaar2024,kim_intentflow_2026,wu_ai_2022}. Importantly, the scaffold was not a fixed prompt template: the options shown were generated and updated from the participant's current request, prior selections and current phase of work. The selected controls then informed the writing interaction for Task~1 only.

\subsubsection{Baseline Interface}

The control condition presented the same conversational AI and editable canvas without the goal specification and task decomposition panels. We selected this naturalistic control because our primary interest was the total effect of adding the complete dynamic scaffold to a familiar AI-assisted writing workflow: a standard chat-and-canvas interface is the workflow into which such a scaffold would ordinarily be introduced and therefore a pragmatically meaningful baseline. It also preserves user agency: control participants could clarify goals, plan and decompose the task through ordinary prompting if they chose to, but the interface did not require or prime those activities.

\subsubsection{Technical implementation.}

Both prototypes were browser-based web applications (React and Node.js). In the scaffold condition, an interface-generation service returned a structured specification of the controls to render from the participant's initial request, which the browser client rendered as interactive components while maintaining the participant's selections across the two phases; a separate service handled chat responses and document drafting using the task brief, conversation history and current control state. The application logged participant messages, scaffold phase, control changes, document generations, canvas interactions and timing.

\subsection{Tasks}

We defined open-ended but bounded knowledge-work tasks that required persuasion, synthesis, audience sensitivity, and explicit coordination with AI. Each task consisted of producing a realistic knowledge-work brief: (1) a city-planning proposal on whether to adopt a 4-day work week for municipal employees; and (2) a consulting brief on how to launch a hybrid-work model for a major client (see Appendix \ref{app:scenariobriefs}). Participants saw the full scenario description at the start of each task and could open a brief task reminder while working. Sessions were self-paced with a maximum completion time of 80 minutes and time on task was recorded for each writing task. To increase motivation and effort, all participants were offered a £20 bonus (or equivalent) for each brief ranked in the top five.

\subsection{Measures}

\subsubsection{Essay quality.}

The primary outcome was essay quality, scored using GPT-5.4 with high reasoning effort against a seven-item analytic rubric, each item rated 1--4: \textit{context and purpose}, \textit{recommendation clarity}, \textit{argument quality}, \textit{evidence credibility}, \textit{feasibility}, \textit{impact} and \textit{organization} (items A--G; \autoref{app:rubric}), and averaged across the seven items. The judge was calibrated against a human rating benchmark as described below~\cite{zheng2023judging,pan-etal-2024-human, LiJudgeChallenges2025,GuJudgeSurvey2024,LiewTan2025,Sessler2025,Xiao2025AES}.

An initial scoring pass on formatted essay text showed sensitivity to layout cues: scaffold-condition essays contained more visible headings and denser section marking, and these features explained additional variance in judge scores beyond the human benchmark. Scoring was therefore conducted on de-formatted essays, stripped of all markup and visual formatting (headings, emphasis, list markers, indentation, rules and repeated whitespace); heading text was retained as ordinary lines, paragraph breaks were normalized, and all lexical content and sentence order were preserved.

\paragraph{Human ratings for calibration and comparison.} A separate sample of Prolific raters (£25 per hour) rated 117 essays (20\% of the 594-essay corpus; selected at random, stratified by task and condition), two raters per essay, as participants had formatted them, on the seven rubric dimensions shared with the LLM judge (items A--G) in a custom interface (\autoref{app:rating-interface}); the two ratings were averaged within essay to form the human benchmark.

\paragraph{Judge calibration.} Each de-formatted essay was scored in five consecutive runs, averaged into the final judge score. The scoring prompt contained the full rubric and 10 human-rated calibration essays (capped at 1,800 characters) carrying their averaged human ratings on A--G; these were excluded from agreement statistics.

\paragraph{Judge comparison.} Because the primary outcome depended on an LLM judge, we compared its scores with the human benchmark on the remaining 107 essays, using the essay-level composites averaged across the seven shared dimensions. The primary agreement statistic was the absolute-agreement intraclass correlation, ICC(A,1), supplemented by the metrics in \autoref{tab:judge-agreement}. Its 95\% confidence interval was obtained from 2,000 author-cluster bootstrap resamples.

\subsubsection{Prompt Behavior.}

To characterize participants' process after scaffold removal, we analyzed the participant-authored prompt sequences from the participant--AI chat logs in Task~2. Each sequence was coded on two dimensions using the codebook in \autoref{app:codebook}: \textit{goal clarity} (GC), which captures how explicitly participants stated the intended outcome, audience, constraints and success criteria (seven indicators), and \textit{task decomposition} (TD), which captures whether they divided the work into concrete steps or components (six indicators). Because goal specification and task decomposition can unfold over several conversational turns, the unit of analysis was the entire participant-authored prompt sequence rather than any single message. All sequences were supplied to coders in a standardized plain-text form with condition labels and assignment metadata removed.

\paragraph{Codebook reliability.} To establish that the codebook could be applied consistently, three of the researchers independently coded 60 of the sequences (20\%), blind to condition, assigning the codebook's ordinal GC and TD ratings (0--3) and marking each of the 13 indicators as present or absent. An LLM coder (GPT-5.4, temperature 0, no extended reasoning) then applied the same codebook to the same 60 sequences, making one request per indicator and one per overall rating, without access to the human codes. Agreement among the human coders, and between the LLM coder and the human coders, is reported in \autoref{tab:rater-reliability}.

\paragraph{Analysis scoring.} For the transfer analysis itself, the LLM coder scored every Task~2 sequence with a graded version of the same 13 indicators: rather than marking each as present or absent, it rated each from 0 (absent) to 3 (well specified), quoting the passages of the participant's text on which the rating rested, in five independent runs. The GC (0--21), TD (0--18) and Combined (0--39) scores are the sums of these indicator ratings, averaged across the five runs. The coder saw the task brief and reminder but no information about condition. Some participants pasted the task wording verbatim as their prompt, so their process could not be meaningfully coded: if any prompt reproduced at least half of the task reminder (13 consecutive words of its 26) or at least half of the scenario description, every indicator in that sequence was set to zero, and the sequence was retained in the analyses.

\subsubsection{Metacognitive Self-Report}

Participants completed 13 pre- and post-task items concerning goal specification, planning, task decomposition, and writing metacognition. Eight items were adapted from the Metacognitive Awareness Inventory~\cite{SchrawDennison1994}, and five drew on the Metacognitive Strategies in Writing questionnaire~\cite{QinZhangXiao2022} (of which four were adapted from the Metacognitive Strategies of Writing Survey~\cite{AlMoqbali2020}). The wording and sources of all items are reported in \autoref{tab:metacognition-items}.

\subsubsection{Subjective Experience Outcomes.}

Participants rated perceived AI helpfulness and control on 1--5 scales. Cognitive load was measured after each task using five NASA-TLX dimensions: mental demand, temporal demand, perceived performance, effort and frustration~\cite{HartStaveland1988}; physical demand was not administered. Items were rated from 1 to 10; the positively worded performance item was reverse-coded as $11-x$ so that higher values indicated greater load, and the five items were averaged with equal weight when all five were observed into an unweighted raw NASA-TLX score. Nine additional study-specific questions about the demands of AI-assisted writing were analyzed at item level (\autoref{tab:cognitive_load_items}) and not included in the overall score; their equal-weight mean, requiring all nine responses, was used as a recall auxiliary.

\subsubsection{Topic Recall Outcomes}

Three days after the writing session, participants were invited to complete a voluntary follow-up survey about the topics of both essays. Among the returners, 20 randomly picked participants received an extra £30 reward through Prolific as incentive. Despite the potential reward being announced at the end of task 2, attrition was high: 159 participants returned to complete the follow-up survey (we address this in Section \ref{subsubsec:recall-missingness}). The analyzed topic-recall outcome was free recall of the writing topic for each of the tasks, in answer to: \textit{``What topic did you write about in your [first/second] essay? If you don’t remember, you can write “Don’t remember”.''} (see \autoref{tab:survey-items} for full survey wording). The open-text responses for each task were coded manually into binary correct/incorrect, scored leniently: a response was credited for each task if it identified the assigned topic or the other task's topic, allowing for confusion about task order. %

\subsubsection{Covariates}\label{subsubsec:covariates}

Unless otherwise stated, analysis models were adjusted for six baseline indicators: whether participants reported using AI for writing, baseline AI-use frequency, self-rated essay-writing skill, task decomposition, AI prompting (preparing appropriate information for the AI), and AI evaluating (reviewing one's judgment before accepting or rejecting an AI recommendation), the latter two taken from the AI metacognition scale of \citet{sidra_generative_2025}. This subset covers writing-specific experience, perceived task competence and relevant metacognitive processes while limiting overlapping item content. Models also included outcome-specific design terms described below.

\subsection{Statistical Analysis}

\subsubsection{Writing quality}

Essay quality for the two tasks was analyzed with linear mixed-effects models including a participant random intercept (for the repeated measures across tasks), fixed effects for task order, treatment condition and their interaction, topic fixed effects for scenario differences, and the common covariate set (\autoref{subsubsec:covariates}). Because the scaffold added an explicit planning stage, Task~1 time on task was partly a mechanism of the intervention, so the primary model omitted it; as a robustness check, we re-estimated the model with log-transformed time on task (seconds) as an additional covariate. The Task~1 and Task~2 contrasts were essentially unchanged.

Rubric item-level contrasts were estimated for each of the seven dimensions separately, in two samples. For the judge across the full 594-essay corpus, each dimension was analyzed with the same mixed-model specification as the composite. For the 107-essay human benchmark, analyzed for both the judge (de-formatted text) and the human raters (formatted text), the composite and dimension-level models used the same fixed effects without a random intercept, estimated by ordinary least squares with standard errors clustered by participant. The benchmark analyses estimate separate between-subjects condition contrasts within Task~1 and Task~2; we do not test their difference. The rubric-level analyses are exploratory and are reported with uncorrected $P$ values; the composite contrasts are the primary tests. Standardized effect sizes for the essay-quality contrasts were computed at the participant level within each task as Hedges' $g$, using the pooled raw standard deviation of the de-formatted essay score; we additionally report the covariate-adjusted between-condition difference divided by the same pooled raw standard deviation.

\subsubsection{Survey Measures}

For each of the 13 metacognitive self-report items, we fitted a separate linear regression with HC3 heteroskedasticity-robust standard errors, estimating the post-task condition contrast adjusted for the corresponding baseline response, first-task topic, baseline AI-use frequency and writing skill, with Benjamini--Hochberg correction across the 13 tests. AI helpfulness and perceived control were analyzed with analogous models. The unweighted five-dimension NASA-TLX score for both tasks was analyzed with a linear mixed model including condition, timepoint and their interaction, with a participant random intercept; simple condition contrasts were derived for Task~1 and Task~2. The supplementary demand items were analyzed separately and not included in the composite.

\subsubsection{Topic Recall}\label{subsubsec:recall-missingness}

A total of 159 participants provided recall outcomes for both tasks. Binary recall outcomes were analyzed using logistic mixed-effects models with a participant random intercept. Fixed effects included scaffold condition, task position (first or second), their interaction, writing scenario, and the six common covariates (\autoref{subsubsec:covariates}). Planned condition contrasts for each task were calculated from the fixed-effect coefficients and their covariance matrix and exponentiated to obtain conditional odds ratios (ORs) for the treatment effect (scaffold versus control in Task~1; previously scaffolded versus control in Task~2).

\paragraph{Accounting for attrition.} Attrition in the follow-up survey was high, though follow-up rates were similar across conditions (49.1\% of scaffolded participants and 49.7\% of controls). We therefore examined the nature of the missing data and assessed sensitivity via multiple imputation \cite{wijesuriya_multiple_2025,white_multiple_2011}.

Treatment assignment did not predict missingness (OR $=1.03$, 95\% CI [0.66, 1.59], $P=.911$; logistic regression on the recall cohort, with one indicator per participant, since both recall outcomes were always jointly observed or jointly missing). To test whether other observed characteristics predicted missingness, we fitted a Firth-penalized logistic regression~\cite{HEINZE2002217} on 26 predictors, including treatment, baseline and outcome survey scores, essay quality scores and other predictors, using the 234 participants with complete predictor data; coefficients are reported uncorrected in \autoref{tab:missingness-firth}. Jointly, observed characteristics did not predict missingness (penalized likelihood-ratio $\chi^2(26)=26.40$, $P=.441$). Two of 26 coefficients had uncorrected $P<.05$, against 1.3 expected under a global null: higher Task~1 essay quality was associated with lower odds of missingness (OR per SD $=0.67$, 95\% CI [0.47, 0.94], $P=.021$) and higher combined Task~2 prompt quality with higher odds (OR per SD $=1.36$, 95\% CI [1.02, 1.80], $P=.033$). The imputation model included essay quality as an auxiliary variable but excluded coded prompt quality.

To assess sensitivity to missing follow-up data, we additionally imputed the missing recall data using multiple imputation by chained equations (MICE) \cite{wijesuriya_multiple_2025,white_multiple_2011}. The imputation model represented the two recall outcomes in separate columns and included condition, first-task scenario, their interaction and the six common baseline covariates. Auxiliaries comprised ten complete-item survey means covering general and writing metacognition, AI metacognition, prompting, workload and AI-writing demand at their administered timepoints; both essay-quality scores; log durations; log-transformed prompt and essay-word counts ($\log(1+x)$); helpfulness and perceived control; and submission, cohort-membership, repeated/missing-form, follow-up-intent and incentive-awareness indicators. 
Specifically, binary variables were imputed by logistic regression and other numeric variables by predictive mean matching with five donors, observed values were preserved, and 50 completed datasets were generated for the recall cohort (100 iterations per chain). The recall model was fitted in each dataset and task-specific log-odds contrasts were pooled with Rubin's rules and $t$ inference.

\subsubsection{Prompt coding}

Prompt-coding analyses were conducted at the prompt-sequence level. To measure Task~2 transfer behavior, we compared previously scaffolded and previously unscaffolded participants on the GC, TD and Combined scores using ordinal logistic regression with the six common covariates (\autoref{subsubsec:covariates}). For the 60-sequence reliability subset, agreement between each pair of coders, human--human and LLM--human, was summarized with quadratic weighted Cohen's $\kappa$. Because the average of three human coders is a more dependable reference than any one of them, we also report agreement between the LLM coder and the mean of the three human ratings as an absolute-agreement intraclass correlation, ICC(A,1), together with the reliability of that mean, ICC(A,3). Confidence intervals were obtained from 2,000 sequence-level bootstrap resamples. Validation of the essay judge is described under Measures.

\section{Results} 

We report the results in the order in which they bear on the study's main argument: essay quality, transfer and process, short-lag topic recall, and subjective experience. Task~1 contrasts compare concurrent scaffold use with the control interface; Task~2 contrasts compare previously scaffolded with previously unscaffolded participants, when neither group had access to the scaffold. Full model summaries are reported in the appendix.

\subsection{The scaffold improved essay quality}

Our primary question was whether the scaffold improved writing quality while it was available and whether any advantage carried into the subsequent unscaffolded task. Essay quality was assessed along seven rubric dimensions (A--G), each on a 1--4 scale, using the de-formatted essay text and calibrated LLM judge described in Methods. We report the overall score before examining the dimensions separately. 

\subsubsection{Overall Essay Quality}

\begin{figure}[htbp]
\centering
\includegraphics[width=0.97\linewidth]{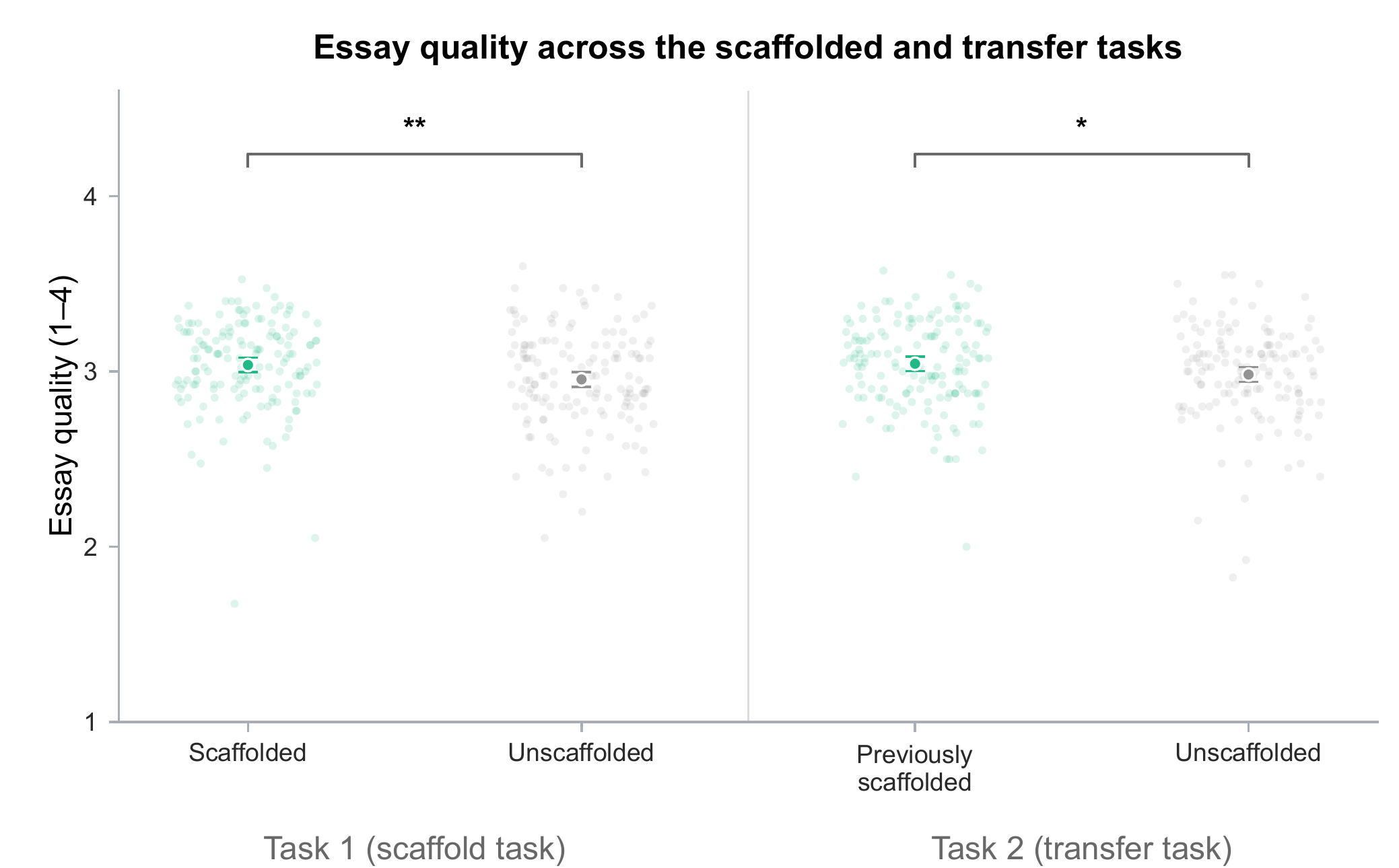}
\caption{Essay quality by condition and task. Scores are means across the seven 1--4 rubric ratings. Faint points show observed essay-level scores; larger markers show condition means with 95\% confidence intervals. Neither group had the scaffold in Task~2. Both planned adjusted contrasts were statistically reliable. $^{**}P<.01$; $^{*}P<.05$.}
\label{fig:essayquality}
\end{figure}

For the primary measure, we computed an overall quality score averaged across the seven rubric dimensions (see \autoref{subsubsec:qualitydimensions} for per-rubric analyses). Scaffolded writers scored higher than controls (mixed-model treatment coefficient $b = 0.089$, 95\% confidence interval (CI) [0.023, 0.155], $P = .008$; composite row of \autoref{tab:rubric-contrasts}; full model in \autoref{tab:essay-models}). The treatment-by-order interaction was negative but not statistically reliable ($b = -0.019$, 95\% CI [$-0.102$, $0.065$], $P = .665$), providing no clear evidence that the condition difference varied between tasks. The Task~2 contrast remained positive ($b = 0.070$, 95\% CI [0.004, 0.136], $P = .037$), where both groups wrote without the scaffold (\autoref{fig:essayquality} shows the raw score distributions). The corresponding raw standardized differences on the 1--4 essay scale were Hedges' $g = 0.29$ in Task~1 and $g = 0.21$ in Task~2; covariate-adjusted differences divided by the same pooled raw standard deviations were 0.28 and 0.23, respectively (\autoref{tab:standardized-effects}).

\FloatBarrier

Because the scaffold introduced an additional planning stage, we asked whether the quality advantage simply reflected time spent. Scaffolded participants spent longer on Task~1 (27.84 versus 16.35 minutes; time ratio 1.56 from a log-duration mixed model with the same fixed effects and participant random intercept as the primary quality model, $P<.001$) but not on Task~2 (10.78 versus 10.62 minutes). Adding log elapsed time left both contrasts unchanged (\autoref{tab:essay-models}), and time on task was not independently associated with quality conditional on treatment ($b=-0.0013$, $P=.951$); the advantage was therefore not accounted for by time on task.

\paragraph{Judge-validation check.} Because the primary quality outcome was LLM-scored, we used the 107-essay human benchmark as a measurement check. Agreement between the essay-level human and judge composites was ICC(A,1) $=0.456$, 95\% CI [0.349, 0.547], with Pearson $r=0.662$ (\autoref{tab:judge-agreement}). More importantly for interpreting the treatment effect, the analysis on the human-rated essays supported the scaffold advantage in both tasks, and in the same 107 essays the judge's contrasts on the seven-dimension composite were smaller than the human raters' (\autoref{tab:rubric-contrasts}); the LLM-based estimate therefore does not appear inflated relative to human judgment of the essays as delivered. %

\subsubsection{Individual Rubric Analysis}\label{subsubsec:qualitydimensions}

The overall score raises a further question: which aspects of writing benefited from the scaffold? We therefore analyzed each of the seven dimensions separately (\autoref{tab:rubric-contrasts}), first for the LLM judge across the full set of 594 essays, using the same mixed model as the composite, and then for both the judge and the human raters within the 107-essay benchmark. The full-corpus analysis provides the broadest picture of the scaffold's contribution, while the benchmark offers a complementary comparison of the two rating approaches on the same essays. These rubric-level analyses are exploratory and use uncorrected $P$ values to characterize the pattern across dimensions. With 52 Task~1 and 55 Task~2 essays, the benchmark yields less precise estimates, but helps assess whether the direction of the composite advantage is shared across rating approaches and which qualities each approach emphasizes. Its contrasts describe the advantage within each task; persistence within individuals is assessed using the full corpus.

\begin{table}[htbp]
\centering
\footnotesize
\setlength{\tabcolsep}{4pt}
\caption{Rubric-level scaffold contrasts, reported as $b$ [95\% CI], $P$.
Full-set models include a participant random intercept, condition $\times$ task,
topic and the common set of six baseline covariates
(\autoref{subsubsec:covariates}); models for the 107-essay benchmark include
the same fixed effects without a random intercept, with participant-clustered
standard errors. Task~1: scaffold minus control; Task~2: previously scaffolded
minus previously unscaffolded. Rubric-level analyses are exploratory;
all $P$ values are uncorrected.}
\label{tab:rubric-contrasts}
\begin{tabular}{@{}lll@{}}
\toprule
Dimension & Task 1 & Task 2 \\
\midrule
\multicolumn{3}{@{}l}{\textit{LLM judge: full set of 594 essays}} \\
Context and purpose
& 0.083 [$-0.007$, 0.173], $P=.071$
& 0.144 [0.054, 0.234], $P=.002$ \\
Recommendation clarity
& 0.057 [$-0.033$, 0.147], $P=.216$
& 0.030 [$-0.060$, 0.120], $P=.514$ \\
Argument quality
& 0.122 [0.040, 0.204], $P=.004$
& 0.078 [$-0.004$, 0.160], $P=.062$ \\
Evidence credibility
& 0.115 [0.026, 0.204], $P=.011$
& 0.039 [$-0.050$, 0.128], $P=.390$ \\
Feasibility
& 0.085 [$-0.008$, 0.177], $P=.074$
& 0.056 [$-0.036$, 0.149], $P=.234$ \\
Impact
& 0.108 [0.018, 0.197], $P=.019$
& 0.094 [0.005, 0.184], $P=.039$ \\
Organization
& 0.052 [$-0.031$, 0.135], $P=.217$
& 0.050 [$-0.033$, 0.133], $P=.239$ \\
Seven-dimension composite (A--G)
& 0.089 [0.023, 0.155], $P=.008$
& 0.070 [0.004, 0.136], $P=.037$ \\
\midrule
\multicolumn{3}{@{}l}{\textit{LLM judge: 107-essay benchmark}} \\
Context and purpose
& 0.048 [$-0.186$, 0.282], $P=.689$
& 0.073 [$-0.171$, 0.318], $P=.556$ \\
Recommendation clarity
& 0.075 [$-0.135$, 0.286], $P=.485$
& 0.081 [$-0.129$, 0.292], $P=.449$ \\
Argument quality
& 0.106 [$-0.060$, 0.273], $P=.211$
& 0.030 [$-0.186$, 0.245], $P=.786$ \\
Evidence credibility
& 0.096 [$-0.156$, 0.349], $P=.454$
& 0.023 [$-0.162$, 0.208], $P=.806$ \\
Feasibility
& 0.004 [$-0.197$, 0.205], $P=.969$
& 0.057 [$-0.189$, 0.302], $P=.651$ \\
Impact
& 0.094 [$-0.123$, 0.312], $P=.396$
& 0.049 [$-0.185$, 0.283], $P=.683$ \\
Organization
& 0.049 [$-0.149$, 0.246], $P=.629$
& 0.151 [$-0.059$, 0.361], $P=.160$ \\
Seven-dimension composite (A--G)
& 0.068 [$-0.076$, 0.211], $P=.357$
& 0.066 [$-0.101$, 0.233], $P=.437$ \\
\midrule
\multicolumn{3}{@{}l}{\textit{Human ratings: 107-essay benchmark}} \\
Context and purpose
& 0.486 [$-0.028$, 0.999], $P=.064$
& 0.482 [0.051, 0.912], $P=.028$ \\
Recommendation clarity
& 0.497 [0.075, 0.920], $P=.021$
& 0.267 [$-0.097$, 0.632], $P=.150$ \\
Argument quality
& 0.032 [$-0.431$, 0.494], $P=.894$
& 0.381 [0.013, 0.749], $P=.043$ \\
Evidence credibility
& $-0.048$ [$-0.556$, 0.460], $P=.853$
& 0.461 [0.031, 0.890], $P=.036$ \\
Feasibility
& 1.198 [0.752, 1.644], $P<.001$
& 1.008 [0.574, 1.442], $P<.001$ \\
Impact
& 0.355 [$-0.245$, 0.955], $P=.247$
& 0.137 [$-0.412$, 0.685], $P=.625$ \\
Organization
& 0.562 [0.155, 0.969], $P=.007$
& 0.293 [$-0.099$, 0.686], $P=.143$ \\
Seven-dimension composite (A--G)
& 0.440 [0.157, 0.724], $P=.002$
& 0.433 [0.120, 0.746], $P=.007$ \\
\bottomrule
\end{tabular}
\end{table}

\paragraph{Rubric-level results across the full set.} Across all 594 essays, the judge's results suggest that the scaffold supported several substantive aspects of writing. In Task~1, advantages were reliable for \textit{Argument quality} ($P=.004$), \textit{Evidence credibility} ($P=.011$) and \textit{Impact} ($P=.019$), suggesting stronger arguments, more credible supporting evidence and better consideration of consequences. Estimates for \textit{Context and purpose} and \textit{Feasibility} were also positive, although they did not reach statistical significance ($P=.071$ and $P=.074$). In Task~2, after the scaffold was removed, advantages remained evident in \textit{Context and purpose} ($P=.002$) and \textit{Impact} ($P=.039$), alongside a positive estimate for \textit{Argument quality} ($P=.062$). \textit{Recommendation clarity} and \textit{Organization} did not differ reliably between conditions in either task. Taken together, this exploratory pattern suggests that the scaffold's contribution extended across several substantive dimensions during use, with evidence of subsequent benefits in how writers framed their briefs and considered their consequences.

\paragraph{Robustness across raters.} The 107-essay benchmark provides a complementary perspective on the scaffold's benefits. Human raters of the formatted essays identified the largest advantage in \textit{Feasibility}, reliable in both tasks ($P<.001$), together with \textit{Organization} ($P=.007$) and \textit{Recommendation clarity} ($P=.021$) in Task~1 and \textit{Context and purpose} ($P=.028$), \textit{Evidence credibility} ($P=.036$) and \textit{Argument quality} ($P=.043$) in Task~2. All seven of their Task~2 estimates were positive, and removing any single dimension left their composite advantage reliable in both tasks, indicating that the human-rated result was not dependent on one dimension alone. The de-formatted judge's benchmark composite estimates also pointed in a positive direction in both tasks (0.068 and 0.066), although they were smaller than the human-rated estimates (0.440 and 0.433). Its dimension-level contrasts were not individually significant in this smaller sample; nevertheless, its largest Task~1 estimates were for \textit{Argument quality}, \textit{Evidence credibility} and \textit{Impact}, the same three dimensions that showed reliable advantages in the full corpus. The two rating approaches therefore agree in the direction of the composite estimates while highlighting different aspects of writing: human ratings particularly emphasize feasibility, and the full-corpus judge results emphasize argument quality, evidence credibility and impact. These exploratory findings suggest that the scaffold's contribution can be visible through complementary perspectives on writing quality.

The \textit{Organization} results bear on concerns that LLM judges favor surface presentation or formatting cues~\cite{zheng2023judging}: the de-formatted judge showed no \textit{Organization} advantage in either task, in the full set or the subset, whereas human raters of the formatted essays did in Task~1, so whatever structural cues survived de-formatting did not drive the judge-rated effect.
\FloatBarrier

\subsection{The scaffold improved later prompting}

The Task~2 quality advantage could reflect transfer of a more deliberate way of specifying work to the AI. To examine this, we compared the prompt sequences participants produced after the scaffold had been removed, coded blind to condition for goal clarity (GC) and task decomposition (TD; Methods). Reliability estimates are reported in \autoref{tab:rater-reliability} and the codebook in \autoref{app:codebook}: the three human coders agreed with one another moderately to substantially, and the LLM coder agreed with their consensus about as well as they agreed with each other on GC and the combined score, and somewhat less closely on TD.

\begin{figure*}[htbp]
\centering
\includegraphics[width=0.98\linewidth]{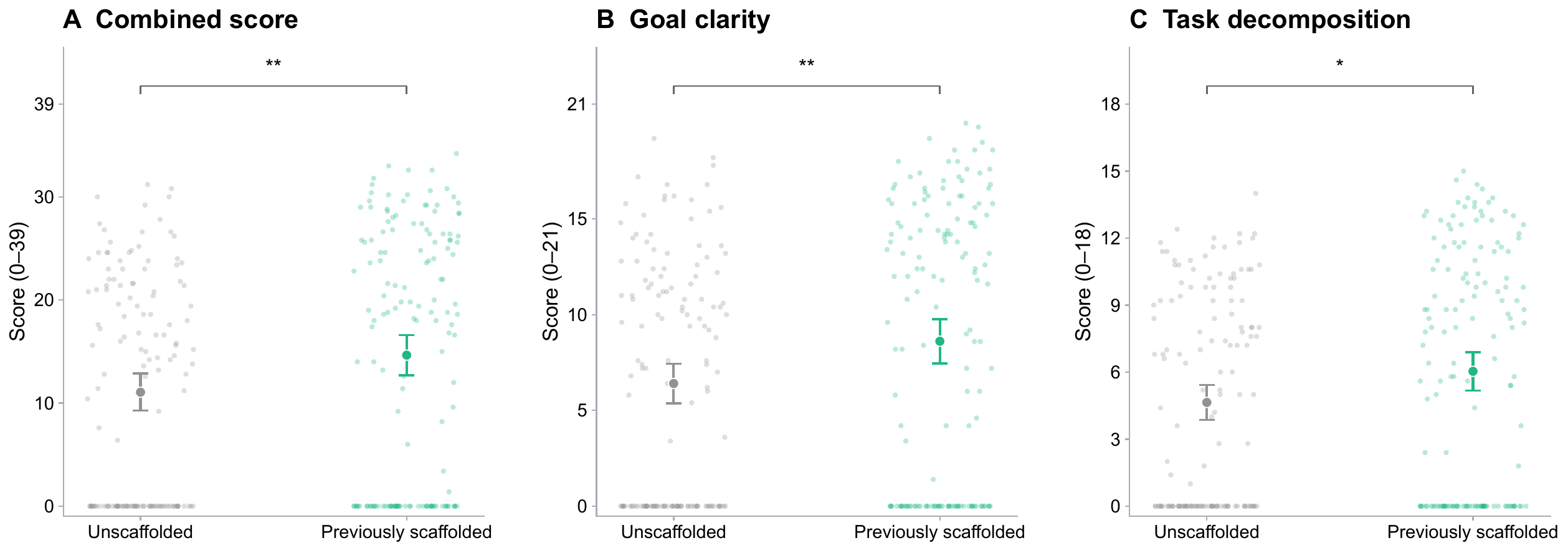}
\caption{Blind coding of prompt behavior in the unscaffolded Task~2. \textbf{A}, Combined goal-clarity and task-decomposition score (0--39); \textbf{B}, goal clarity (seven indicators; 0--21); \textbf{C}, task decomposition (six indicators; 0--18). Faint points: participants' mean scores across five LLM coder runs ($N=297$); circles: condition means with 95\% participant-bootstrap CIs (10,000 resamples). Brackets: ordinal regressions with the six common covariates (\autoref{subsubsec:covariates}); Combined $P=.0067$, GC $P=.0036$, TD $P=.0431$.} \label{fig:promptcoding}\end{figure*}

Sequences that reproduced the task wording verbatim were scored zero (see Methods): 56 of 153 previously scaffolded participants (36.6\%) and 63 of 144 previously unscaffolded participants (43.8\%), visible as the concentration at zero in \autoref{fig:promptcoding}. By itself, this indicates that a substantial proportion of participants offloaded the task to the AI despite being incentivized for high performance.

Nevertheless, previously scaffolded participants scored reliably higher than previously unscaffolded participants on the combined GC+TD score (14.43 versus 11.07 of a possible 39) and on GC and TD separately (ordinal logistic regressions; \autoref{tab:prompt-effects}; score distributions in \autoref{fig:promptcoding}). 

\begin{table}[htbp]
\centering
\footnotesize
\caption{Task~2 prompt behavior by condition: descriptive statistics, standardized effect sizes and ordinal logistic regression estimates with the six common covariates (Section~\ref{subsubsec:covariates}). OR is the treatment effect of prior scaffolding versus control on the odds of a higher score; $N=297$.}
\label{tab:prompt-effects}
\setlength{\tabcolsep}{3pt}
\begin{tabular}{@{}lrrrrrrlr@{}}
\toprule
Measure & \shortstack{Prior-scaffold\\$M$ (SD)}
& \shortstack{Control\\$M$ (SD)}
& \shortstack{Mean\\difference}
& Cohen's $d$
& \shortstack{Rank-biserial\\$r$}
& \shortstack{Probability of\\superiority}
& OR [95\% CI]
& $P$ \\
\midrule
Goal clarity & 8.49 (7.25) & 6.43 (6.25) & 2.067 & 0.30 & 0.19 & 0.60 & 1.862 [1.225, 2.828] & .0036 \\
Task decomposition & 5.94 (5.37) & 4.65 (4.66) & 1.291 & 0.26 & 0.14 & 0.57 & 1.537 [1.013, 2.332] & .0431 \\
Combined GC+TD & 14.43 (12.29) & 11.07 (10.67) & 3.358 & 0.29 & 0.18 & 0.59 & 1.779 [1.173, 2.698] & .0067 \\
\bottomrule
\end{tabular}
\end{table}

\subsection{The scaffold was associated with improved short-lag topic recall}\label{subsec:recall-results}

Memory is a key requirement of cognitive engagement with work. We asked participants to complete a voluntary follow-up survey approximately three days after their session to test recall of their writing topics. Given this delay, attrition was high: 159 participants supplied scored follow-up responses for both tasks. As a robustness check on the complete-case analysis, we therefore also conducted a multiple imputation analysis. 

\subsubsection{Complete-case analysis}

\begin{figure}[htbp]
\centering
\includegraphics[width=0.80\linewidth]{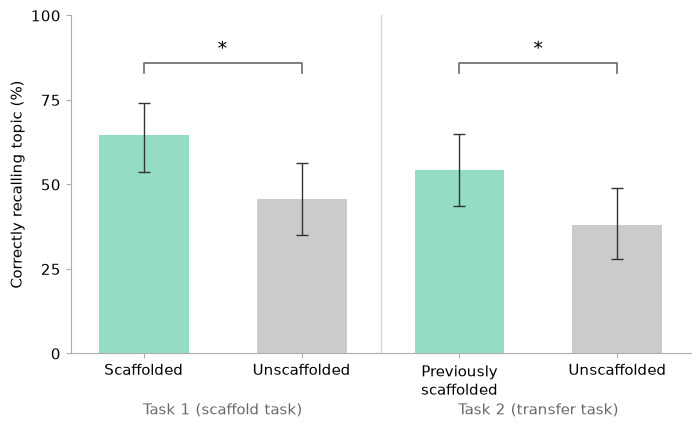}
\caption{Correct topic recall by condition and task. Bars show observed recall percentages with 95\% Wilson confidence intervals. Asterisks indicate adjusted logistic mixed-model contrasts (Task~1, $P=.026$; Task~2, $P=.046$).}
\label{fig:recall}
\end{figure}

Observed recall percentages in the complete-case analysis were 64.6\% for scaffolded participants versus 45.6\% for controls in Task~1, and 54.4\% for previously scaffolded participants versus 38.0\% for controls in Task~2. In the complete-case analysis (model in \autoref{subsubsec:recall-missingness}), scaffolded participants had better topic recall than controls for Task~1 (treatment-effect OR $=5.07$, 95\% CI [1.21, 21.25], $P=.026$) and Task~2 (OR $=4.24$, 95\% CI [1.02, 17.57], $P=.046$; \autoref{tab:topic-recall-models}; observed rates in \autoref{fig:recall}).

\subsubsection{Multiple imputation analysis}

After multiple imputation for the recall cohort (\autoref{subsubsec:recall-missingness}), the scaffold effect remained positive for Task~1 (OR $=3.24$, 95\% CI [1.11, 9.49], $P=.032$) and Task~2 (OR $=3.06$, 95\% CI [1.04, 9.02], $P=.042$). The positive pattern therefore extended to recall of the task completed after scaffold removal.

\FloatBarrier

\subsection{The scaffold had mixed effects on subjective perceptions}

Objective performance benefits need not correspond to an easier or more positively experienced interaction. We therefore examined whether the scaffold changed participants' metacognitive self-reports, perceived helpfulness and control, and cognitive load.

\subsubsection{Metacognitive self-report}

Of the 13 items (Methods), the scaffold selectively increased participants' reports that they broke the task into smaller steps: in the baseline-adjusted regression, this item was higher in the scaffold condition at post-task ($b = 0.753$, 95\% CI [0.546, 0.959], $P_{\mathrm{BH}} < .001$). The staged composing-process item showed a positive association that did not survive correction ($b = 0.225$, 95\% CI [0.010, 0.440], $P_{\mathrm{BH}} = .233$); no other item-level condition effect was reliable after correction (\autoref{tab:metacognition-items}).

\subsubsection{Helpfulness and perceived control}

Scaffolded participants rated the AI as \textit{less} helpful than unscaffolded participants ($b=-0.342$, 95\% CI [$-0.510$, $-0.174$], $P<.001$) and reported \textit{lower} perceived control ($b=-0.251$, 95\% CI [$-0.486$, $-0.017$], $P=.036$; \autoref{tab:subjective-summary}). Both subjective ratings therefore moved in the opposite direction from the objective Task~1 quality effect.

\subsubsection{Cognitive load}

The NASA-TLX summary score (Methods; \autoref{tab:cognitive_load_items}) showed a transient increase in load: during Task~1, scaffolded participants reported higher load than controls ($M=4.71$ versus $3.96$; $b=0.745$, 95\% CI [0.381, 1.110], $P<.001$). During Task~2, when neither group received the scaffold, the corresponding difference was smaller and not statistically reliable ($M=4.08$ versus $3.93$; $b=0.147$, 95\% CI [$-0.218$, 0.512], $P=.429$). The condition-by-task interaction was reliable ($b=0.598$, 95\% CI [0.270, 0.926], $P<.001$), localizing the additional load to scaffold use.

\begin{table*}[t]
\centering
\caption{Item-level cognitive-load outcomes during the scaffolded and subsequent unscaffolded tasks}
\label{tab:cognitive_load_items}
\scriptsize
\setlength{\tabcolsep}{2.5pt}
\renewcommand{\arraystretch}{1.12}
\begin{threeparttable}
\begin{tabularx}{\textwidth}{@{}>{\raggedright\arraybackslash}X cccccccc@{}}
\toprule
& \multicolumn{4}{c}{Task 1: scaffold active}
& \multicolumn{4}{c}{Task 2: no scaffold} \\
\cmidrule(lr){2-5}\cmidrule(lr){6-9}
Question
& Control $M$
& Scaffold $M$
& $b$ [95\% CI]
& $P_{\mathrm{BH}}$
& Control $M$
& Prior-scaffold $M$
& $b$ [95\% CI]
& $P_{\mathrm{BH}}$ \\
\midrule

```latex
How mentally demanding was the essay writing task? & 4.88 & 5.40 & $0.526$ [$-0.036$, $1.087$] & .087 & 4.86 & 4.73 & $-0.154$ [$-0.730$, $0.421$] & .906 \\
How hurried or rushed was the pace of the task? & 4.20 & 5.23 & $0.983$ [$0.359$, $1.607$] & \textbf{.005} & 3.79 & 4.06 & $0.215$ [$-0.353$, $0.782$] & .906 \\
How hard did you have to work to accomplish your level of performance? & 5.44 & 5.85 & $0.441$ [$-0.115$, $0.997$] & .130 & 4.91 & 5.12 & $0.215$ [$-0.380$, $0.811$] & .906 \\
How insecure, discouraged, irritated, stressed, or annoyed were you? & 2.59 & 3.59 & $0.960$ [$0.425$, $1.495$] & \textbf{.002} & 2.85 & 3.04 & $0.146$ [$-0.373$, $0.665$] & .906 \\
How mentally demanding was the essay topic itself to understand and write about? & 4.55 & 4.83 & $0.272$ [$-0.322$, $0.867$] & .369 & 4.23 & 4.34 & $0.062$ [$-0.523$, $0.647$] & .906 \\
How much mental effort did developing coherent arguments for this essay demand? & 4.91 & 5.38 & $0.496$ [$-0.062$, $1.055$] & .096 & 5.01 & 4.88 & $-0.119$ [$-0.678$, $0.440$] & .906 \\
How much mental burden did managing the AI tool add for writing the essay? & 3.34 & 4.27 & $0.936$ [$0.377$, $1.495$] & \textbf{.003} & 3.37 & 3.62 & $0.225$ [$-0.321$, $0.770$] & .906 \\
How mentally demanding was prompting the AI? & 3.37 & 4.40 & $1.029$ [$0.489$, $1.568$] & \textbf{.001} & 3.64 & 3.83 & $0.184$ [$-0.347$, $0.715$] & .906 \\
Did you have to think hard about how to structure your essay effectively? & 4.76 & 5.36 & $0.605$ [$-0.004$, $1.215$] & .074 & 4.86 & 4.70 & $-0.175$ [$-0.779$, $0.430$] & .906 \\
Did integrating AI suggestions require careful mental processing? & 4.76 & 5.30 & $0.583$ [$0.024$, $1.142$] & .067 & 4.76 & 4.68 & $-0.074$ [$-0.662$, $0.515$] & .906 \\
Did you struggle to keep track of all the ideas you wanted to include? & 3.30 & 4.36 & $1.031$ [$0.483$, $1.579$] & \textbf{.001} & 3.20 & 3.61 & $0.381$ [$-0.130$, $0.892$] & .906 \\
Was it mentally challenging to maintain consistency throughout the essay? & 3.83 & 4.53 & $0.714$ [$0.168$, $1.261$] & \textbf{.019} & 3.72 & 3.76 & $0.015$ [$-0.526$, $0.556$] & .957 \\
Was it difficult to maintain focus throughout the task? & 2.78 & 3.38 & $0.621$ [$0.153$, $1.089$] & \textbf{.019} & 2.87 & 2.96 & $0.092$ [$-0.389$, $0.572$] & .906 \\

\bottomrule
\end{tabularx}
\end{threeparttable}
\end{table*}

\begin{figure*}[htbp]
\centering
\includegraphics[width=0.99\linewidth]{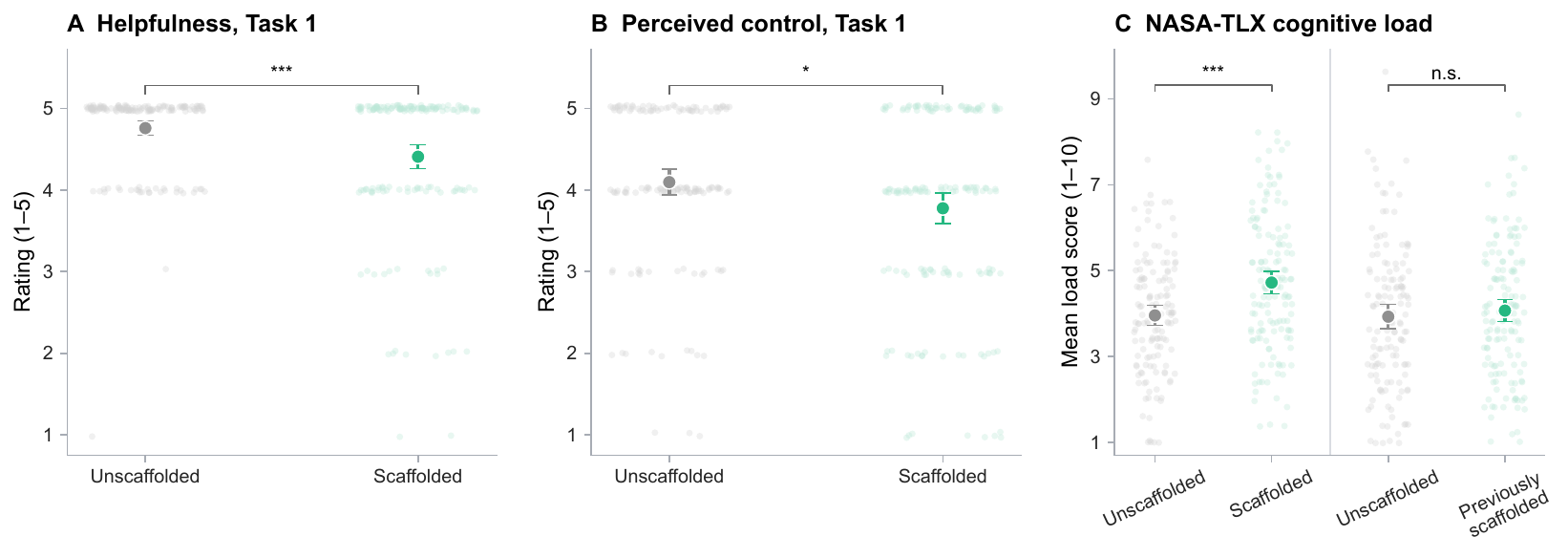}
\caption{Subjective experience by condition. \textbf{A}, AI helpfulness and \textbf{B}, perceived control after Task~1; \textbf{C}, unweighted NASA-TLX load after Tasks~1 and 2 (higher, more load). Faint points: participant ratings with slight vertical jitter (up to $\pm 0.04$ scale units); markers: condition means with 95\% confidence intervals from the unjittered data. $^{***}P<.001$; $^{*}P<.05$; n.s., not significant.}
\label{fig:subjective}
\end{figure*}

The item-level models localized this transient increase (\autoref{tab:cognitive_load_items}; \autoref{fig:subjective}). During scaffold use, 7 of the 13 items remained reliable after correction, with the largest elevations for prompting demand, difficulty tracking ideas, the burden of managing the AI, stress or frustration, and rushed pace, and smaller elevations for maintaining consistency and focus. None of the 13 items differed reliably during the subsequent unscaffolded task; the additional burden was concentrated in coordinating and managing the scaffolded interaction while it was active.

\FloatBarrier

\section{Discussion}\label{sec:discussion}

As generative AI shifts knowledge work from production toward planning and evaluation, this study found that dynamic metacognitive scaffolds improved written output quality, both while the scaffold was available and after it was removed. Participants who had used the scaffold also later wrote prompts with greater goal clarity and task decomposition on their own. We also found evidence for improvement in topic recall three days later. However, even as their written output and prompting improved, scaffolded participants rated the system as less helpful, felt less in control, and reported greater cognitive load while the scaffold was active. Our discussion focuses on the implications of these finding for different levels of knowledge work, the divergence between subjective and objective measures, designing productive friction with visible value into AI assistance, and implications as agentic autonomy increases. Finally, our results have important limitations and present opportunities for future research.

\subsection{Scaffolding different levels of knowledge work}

Our results suggest that metacognitive scaffolding may matter at three different levels of knowledge work: how people approach a task, how they work with the AI on it, and how well they understand the work itself. 

\subsubsection{How people approach the task}

The first level includes \textit{engaging} in metacognitive processes and becoming \textit{aware} of them. The scaffold improved output quality across tasks, and scaffolded participants engaged in more goal specification and task decomposition in their prompts. This latter aspect is where the scaffold had its most reliable effect, with participants both \textit{reporting} more task structuring and actively \textit{doing} so after scaffold removal. They did not, however, report setting clearer goals. Clarifying goals may need more than one supported episode, or a different kind of support, before people both do it \textit{and notice that they have improved}~\cite{Tankelevitch2024,SchrawMoshman1995,Zimmerman2000}; we return to this divergence in \autoref{subsec:friction}. %

\subsubsection{How people work with the AI}

The second level is how people learn to direct AI systems. We have some evidence of short-lag transfer~\cite{BarnettCeci2002}: participants who had used the scaffold went on to write better \textit{essays} and clearer, better-structured \textit{prompts} on their own. Scaffold removal was part of our experimental design, but we are not suggesting that metacognitive support must eventually be removed. Such support, however, cannot be guaranteed in diverse knowledge work contexts. How well someone directs an AI system is much of what separates people who get value from these tools from those who do not, and \textit{that skill is unevenly distributed}~\cite{sarkar-s_aiagents_2026,Kazemitabaar2024}. A brief scaffold that leads people to prompt more deliberately is a cheap way to build a scarce skill, and one that \textit{may} follow the user from one tool to the next, whether scaffolds are available or not. 

\subsubsection{How people understand and engage with the work itself}

The third level is that the effort of goal specification and task decomposition, and the sharper prompting it brings, may help people better understand how tasks fit into larger outcomes and the nature of their work overall \cite{tankelevitch2026goals,stankowski_design_2026}. This third level arguably matters the most to the future of work. %

We suggest that the \textit{initiating} role of goal specification and task decomposition in AI-assisted workflows is key to this, aligning with prior work. In ~\citet{xu_productive-v-reflective_2025}, when AI asked questions, compared to simply generating outputs for review, participants spent more time generating and revising, yet also felt more confident in the process. Compared to no AI support, they also reflected more on the framing and underlying problems, and produced more practical solutions: hallmarks of cognitive engagement that, as in our study, seem to follow from \textit{structuring the task}. Likewise, when time permits, independently \textit{initiating} work before using AI leads to better critical thinking outcomes \cite{zhi_investigating_2026}. 

Memory of work materials is key to understanding the nature of one's work. Prior research has found AI assistance to be detrimental \cite{Bastani2025,Zindulka2026MemoryGap,zhi_investigating_2026,chen_more_2025,kreijkes_effects_2025}: one week after generating and elaborating ideas with and without AI,~\citet{Zindulka2026MemoryGap}'s participants misattributed sources significantly more, most of all in mixed human--AI workflows. Our scaffolded participants also engaged in mixed-human-AI workflows, yet were more likely to recall the topics of their writing three days later. Our comparison is between two AI-assisted conditions, so it does not contradict evidence that AI impairs retention relative to unaided work. Rather, it suggests that metacognitive scaffolds may mitigate this impact relative to default chat interfaces. Memory, and cognitive engagement more broadly, is also richer than topic recall, and future work should explore this. Finally, we interpret our findings cautiously: the contrasts were significant across analyses but estimated imprecisely, and rest on the half of the cohort who returned (see Limitations in Section \ref{sec:limitations}).

\subsection{Divergence Between Subjective and Objective Measures}\label{subsec:divergence}
A tension in our results is that scaffolded participants performed better yet rated the AI as less helpful. This is not unexpected \cite{Tankelevitch2024}, as the scaffolds required participants to analyze the task and make decisions, demanding cognitive effort and time. This echoes the concept of `desirable difficulties' in learning sciences, where certain effortful engagement can feel aversive, yet lead to better learning outcomes \cite{bjork2011making}. For example, active learning that cognitively engages students improves objective learning yet is rated lower by students who \textit{subjectively} experience poorer learning \cite{deslauriers_measuring_2019}. Conversely, fluent instructors can increase interest and perceptions of learning without affecting objective performance \cite{carpenter2020effects}. However, not all difficulties are desirable: designing for \textit{productive} friction is key as we discuss in Section \ref{subsec:friction}.    

A similar dissociation appeared in the subjective and objective measures of metacognition: participants reported breaking the task into smaller steps but not setting clearer goals, whereas their prompts in the unscaffolded task showed gains in \textit{both}. One reading is that the two processes differ in how visible they are to those performing them. Decomposition leaves inspectable traces, such as a list of parts, whereas a clearer goal may simply feel like knowing what one wants, without any awareness that one now specifies it more explicitly~\cite{NelsonNarens1990,Efklides2008, Tankelevitch2024}; self-report registers the process that generates such cues. 

More broadly, subjective and objective measures of AI-assisted work can diverge, which in itself reflects a limitation of metacognition. The pattern recurs across the literature: self-rated critical thinking varied little across conditions whose measured performance differed substantially~\cite{zhi_investigating_2026}; users' estimates of their own contribution shifted once they were shown goal-level traces of their conversations~\cite{kim_i_2026}; the cognitive forcing function that most reduced over-reliance on AI-generated plans was not the one participants found most helpful~\cite{ghosh_experimental_2026}; and experienced developers believed AI had sped them up when it had slowed them down~\cite{Becker2025Slowdown} (see also~\cite{Fernandes2026NoneTheWiser,Bucinca2021}). Offloading feels helpful precisely because it removes the effort that constitutes engagement~\cite{Lee2025,Draxler2024}; in co-writing research, support that improves the product can still feel less satisfying or less like one's own~\cite{Dhillon2024,Li2024,Draxler2024,Hwang2025,joshi2024writing,Heung2025,KIRK2025114984}; and dynamic controls demand inspection and choice even when they improve steering~\cite{drosos_dynamic_2025}. This suggests that studies evaluating AI's impact on cognition, including system evaluations, should include objective measures of performance and cognitive processes \cite{zhi_investigating_2026, zhicombining, tankelevitch_understanding_2025}. 

\subsection{Designing productive friction with visible value into AI assistance}\label{subsec:friction}

A tool that feels less helpful is likely to be declined when it is optional: its benefits arrive only once the task is done, and often only in comparison with others' work, which may never happen. Our data show how strong the pull toward the low-effort path is: despite a bonus for the best briefs, 119 of the 297 participants (40\%) produced Task~2 prompt sequences that met the task-text copying criterion (\autoref{fig:promptcoding}), delegating not just production but the specification of the work itself. This may partly reflect our cohort, but it is also overreliance on AI in its most complete form~\cite{Bucinca2021,ghosh_experimental_2026,shaw_thinkingfast_2026}.The design challenge is to \textit{treat friction as a design material} rather than a cost to be minimized~\cite{zeya_2024_positivefriction,gould_2021_designedfriction,sheahan_2024_designingwithfriction, scott_what_2025}, extending a long line of HCI research from microboundaries that interrupt automatic interaction to seamful designs that keep a system's workings in view~\cite{Cox2016Microboundaries}. 

Not all friction is productive, however. Although we designed our scaffold to be supportive, participants' open-ended comments after Task~1, which we did not analyze systematically, are suggestive. The clearest theme was the cost of specifying everything up front ``in a short time to meet requirements''. This aligns with \citet{zhi_investigating_2026}: the cognitive benefits of independently structuring a task in an AI-assisted workflow requires sufficient time. A related point was the need for ``more time to learn how to use it effectively'', which also affected participants ``ownership of the writing''. A second theme was that choosing between goals itself demands expertise, for example ``prioritising being persuasive or using emotional appeal''--this is partly a limitation of the simulated nature of our task. Finally, another concern was the gap between control over planning and control over revision, such as simple document-wide corrections and ways to ``lock'' sentences. One participant valued the planning support but desired “options to refine what I had done first time”. These comments point to scaffolds that stage specification across the draft rather than front-loading it, that explain the choices they ask for, and that pair metacognitive support for planning with support for revision \cite{kazemitabaar_improving_2024,kim_intentflow_2026}.

More generally, showing the seams of breaking the task down should be part of producing work that one can stand behind and will remember, rather than an obstacle the AI has imposed~\cite{Li2024,he_rebalancing_2023,KIRK2025114984}. This requires making the value legible at three points where it is easy to miss: (a) in the moment, so that instructing and steering the AI carries its own payoff, experienced as \textit{thinking by doing} rather than as friction before the `real work' begins; (b) across the workflow, so that the structure a scaffold imposes is seen to pay off as work moves from one phase to the next rather than registering only as an upfront cost; and (c) in the outcome, so that people can recognize how a result was achieved and feed that recognition into the next piece of work.

\subsection{Supporting Human Metacognition as Agentic Autonomy Grows}\label{subsec:frontier}

These findings should be read against a rapidly moving frontier. It is tempting to assume that more capable agentic systems will simply absorb goal-setting and decomposition. But as autonomous task horizons lengthen~\cite{Kwa2025TimeHorizon}, a single instruction sets off longer chains of action, and human involvement shrinks to fewer, higher-stakes moments of specifying and verifying what the system produces~\cite{sarkar-s_aiagents_2026,passi_agentic_2025,dhanorkar_human_2026}. What most remains with the person, and what they are most anxious to keep, is the understanding and direction of the work itself~\cite{ShaoWorkBank2025}. The socio-technical problem is thus to design for the regulation of work as much as its generation: for \textit{the disposition that an AI system can cultivate in a person}, with scaffolds that fade as the relevant skills are internalized~\cite{MOLENAAR2022100070, Tankelevitch2024}. Here that means specifying goals, breaking work into checkable parts, \textit{and eliciting both} from whatever system is at hand, so that tools leave workers more capable as and when AI systems change or are unavailable. 

\subsection{Limitations}\label{sec:limitations} 
A key limitation is the LLM judge. Firstly, there is a possible bias toward structure rather than content. However, the judge found a smaller treatment advantage than human raters who saw the formatted essays (\autoref{tab:rubric-contrasts}), providing little evidence that structure inflated the primary estimate. The 107-essay benchmark adds a useful comparison of rating approaches on the same essays, with human ratings supporting an advantage in both tasks and the judge's composite estimates pointing in the same direction. Its smaller size limits precision, and its between-subjects comparisons address the advantage within each task; evidence concerning persistence within individuals comes from the full-corpus analysis. Secondly, the judge's `conservatism' is defined relative to human-perceived quality, which itself incorporates presentation, so we have no `structure-free' benchmark. Thirdly, agreement with human ratings was moreover moderate rather than strong, and the judge compressed the human score range (calibration slope $=0.375$; \autoref{tab:judge-agreement}). The essay-quality result should therefore be read together with the prompting, metacognitive self-report and recall evidence, which do not depend on the judge and converge on the same conclusion.

Another limitation concerns our recall analysis. Both contrasts underlying the recall effect were positive across analyses, but the confidence intervals were wide (\autoref{tab:topic-recall-models}), so the data are compatible with anything from a small to a large effect. Recall was also scored under a lenient rule that credited any correct topic identification, a deliberately low bar that leaves open whether the effect extends to deeper understanding or retention of essay content. Follow-up attrition was substantial (50.6\% of participants in the recall cohort had no scored recall outcome). Although attrition is expected in an online sample, it introduces potential for sample biases that limit interpretation and generalizability. Treatment did not predict missingness, whereas other observed factors did, and the complete-case and multiple-imputation analyses agree, which limits the risk of bias in the treatment contrast. However, we can only rule out dependence on what we measured. Finally, topic recall is a narrow index of cognitive engagement, which also involves critical thinking about material~\cite{zhi_investigating_2026,Lee2025} and communicating about work with collaborators~\cite{riedl_cognitive_2026}; future work should use tasks that require evaluating and integrating source material~\cite{zhi_investigating_2026} and communicating outputs.  

Our experimental design has two limitations. Since previous research already demonstrates the value of dynamic over static scaffolds~\cite{drosos_dynamic_2025}, and we prioritized ecological validity in our baseline, we did not have a static scaffold condition. It is possible that the cognitive load effects could be partly mitigated with a static scaffold, particularly longitudinally. Second, our study tests the \textit{combined} impact of a goal specification and task decomposition scaffold, rather than the distinct interface components. We opted for this as both aspects are theoretically interwined and function synergistically ~\cite{Tankelevitch2024,kim_intentflow_2026}.

Finally, the Task~2 findings of improved prompting and output quality show short-lag transfer to a subsequent unscaffolded task, not long-term persistence across weeks or months. The study used one-shot online writing tasks with U.S.-based Prolific adults rather than embedded classroom or workplace deployments, and could not separate effects by experience. Future work should test longitudinal effects in more ecologically valid settings; whether scaffolds close the gap between strong and weak planners or lift workers equally; other domains, such as data science and engineering, that could equally benefit from metacognitive support~\cite{Tankelevitch2024}; and, as agentic workflows gain popularity, long-horizon multi-step tasks, with completion rates and token efficiency as additional outcomes.        

\section{Conclusion}\label{sec:conclusion}

We tested whether a dynamic metacognitive scaffold for goal specification and task decomposition improved AI-assisted writing compared with a standard chat interface. Despite clear benefits to writing and prompting, the scaffold was perceived as imposing additional effort and reduced perceived helpfulness. Designers should focus not only on optimizing ease and output, but also on users’ capacity to calibrate their work practices and outcomes with and without AI. When working with AI, there is a need to design for the visible value of productive friction, to bring AI the benefit of human judgment while ensuring meaningful human contribution.

\FloatBarrier\clearpage	\bibliographystyle{ACM-Reference-Format}
\bibliography{references,references_added,rubric_refs_added,as-refs}

\appendix\clearpage\onecolumn
\setcounter{table}{0}\renewcommand{\thetable}{A\arabic{table}}
\setcounter{figure}{0}\renewcommand{\thefigure}{A\arabic{figure}}
\noindent\textbf{Appendix guide.} The appendix follows the order of the paper. Appendices~\ref{app:probe}--\ref{app:survey-items} document the materials and instruments: the dynamic scaffold interface and scenario briefs, the essay-scoring rubric and human-rating interface, the prompt-coding codebook with coder agreement, and all survey items.  Appendices~\ref{app:res-quality}--\ref{app:res-subjective} report supplementary model summaries in the order of the Results: essay quality, topic recall and follow-up missingness, and subjective outcomes.\medskip

\section{Research probe and tasks}\label{app:probe}

\subsection{Dynamic scaffold interface}\label{app:interface}

The scaffold used during Task~1 was implemented as a two-phase dynamic interface inspired by Dynamic Prompt Middleware / Dynamic Prompt Refinement Controls~\cite{drosos_dynamic_2025}. Rather than exposing all controls at once, the interface staged the metacognitive support. This sequencing was intended to mirror a simple logic of knowledge work: first make the goal explicit, then decide how to break the work down.

\textbf{Goal Clarity phase.} The first phase translated an initially underspecified request into an explicit rhetorical problem. Participants were asked what message they wanted to convey, who the target reader was, what outcome they wanted in that reader, what the primary writing objective was and how deep or comprehensive the response should be. In theoretical terms, this phase externalized core goal parameters that are often implicit in AI prompting. In design terms, it created inspectable fields that the participant could edit before substantial text generation began.

\begin{figure}[htbp]
\centering
\includegraphics[width=0.95\linewidth]{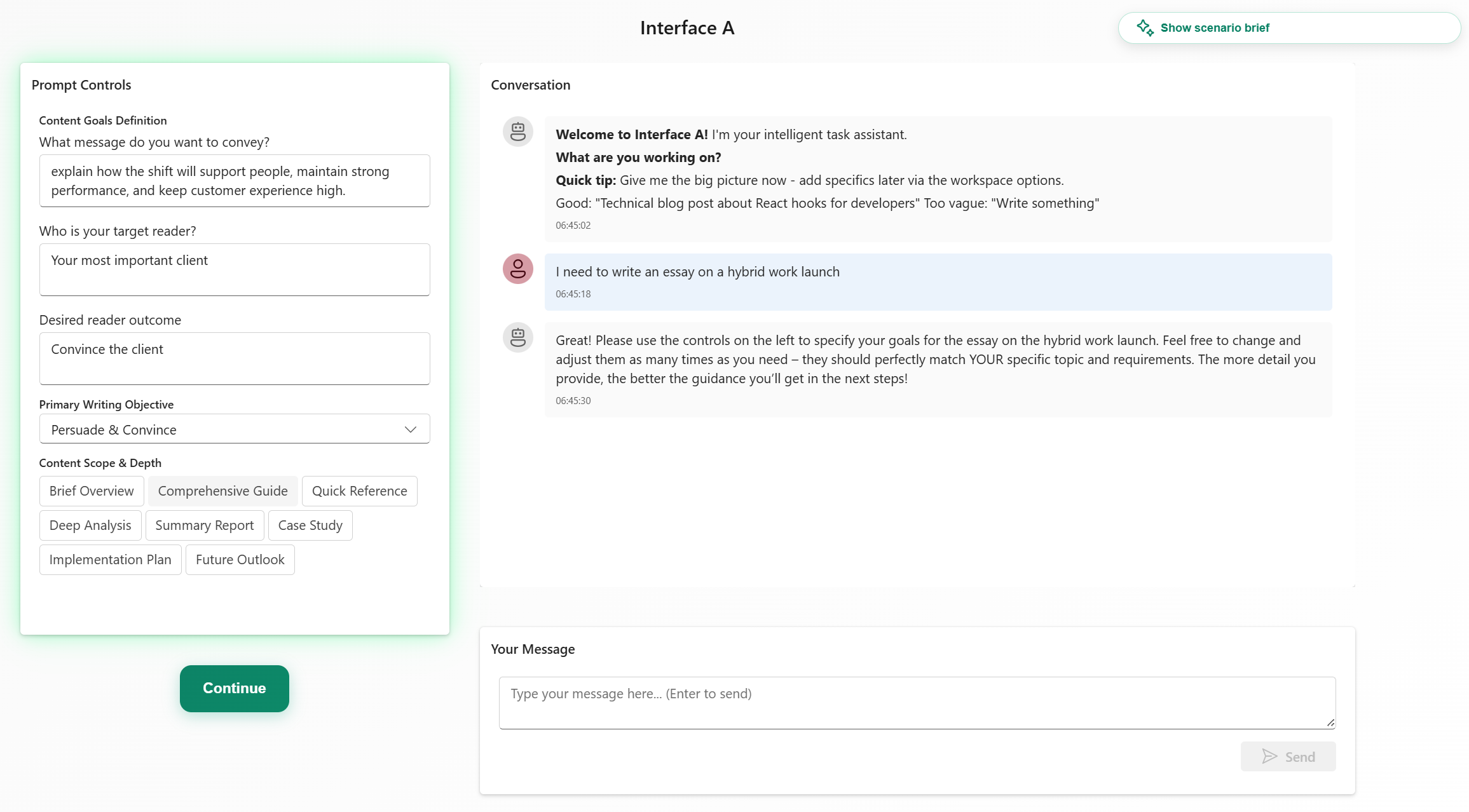}
\caption{Goal Clarity interface. After the participant entered an initial request, the system rendered editable controls for message, audience, intended reader outcome, primary writing objective and desired scope/depth. This phase operationalized goal specification as an explicit, inspectable interface step rather than leaving it embedded only in free-form prompting.}
\label{fig:appendix_goalclarity}
\end{figure}

\textbf{Task Decomposition phase.} After the participant specified the goal-clarity controls, the interface rendered a second panel tailored to the current scenario and the participant's prior selections. In this phase, the system generated task-specific planning controls, including candidate analytical approaches, a selectable essay structure and a field for critical considerations. Participants could accept these controls, revise them directly or request alternative controls through a natural-language workspace box. This phase operationalized decomposition as a structured but editable breakdown of the task into manageable analytical and rhetorical components.

\begin{figure}[htbp]
\centering
\includegraphics[width=0.95\linewidth]{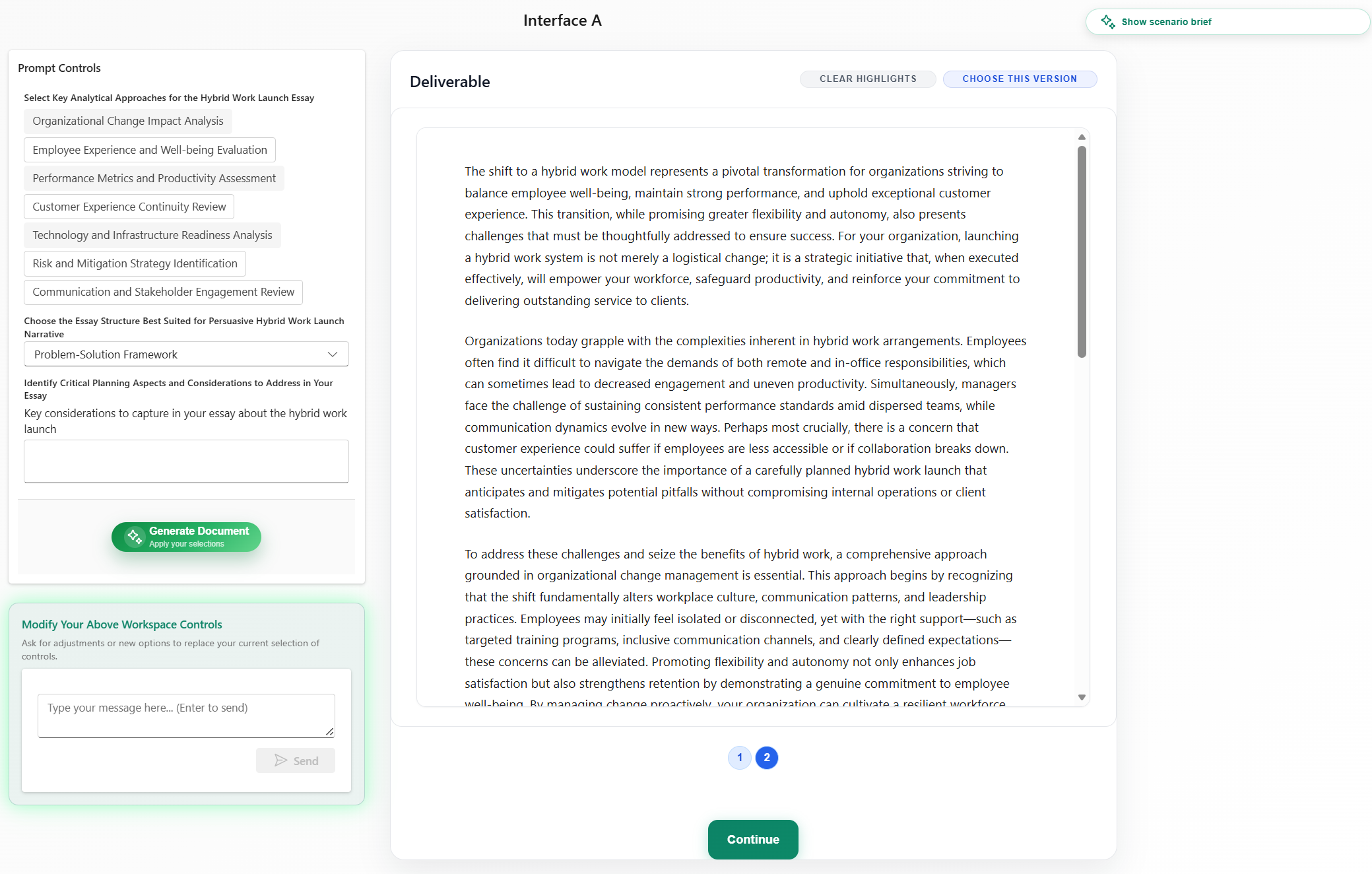}
\caption{Task Decomposition interface. Once goal-clarity information was specified, the system generated task-specific planning controls for analytical approaches, essay structure and critical considerations. A dedicated natural-language box allowed participants to ask for adjustments or replacement controls, making the decomposition adaptive to their evolving intent.}
\label{fig:appendix_taskdecomp}
\end{figure}
\FloatBarrier

Together, these two phases instantiated the central design claim of the intervention: metacognitive support for AI-assisted writing can be rendered as dynamic interface controls that respond to the user's current request and stage of work, rather than as a single static checklist. Goal clarity was intended to constrain the target, and task decomposition was intended to organize execution and evaluation. The interface therefore scaffolded not just what to ask the model, but how to represent and manage the task itself.

\clearpage

\FloatBarrier

\subsection{Scenario briefs}\label{app:scenariobriefs}

\begin{figure}[htbp]
\centering
\includegraphics[width=0.74\linewidth]{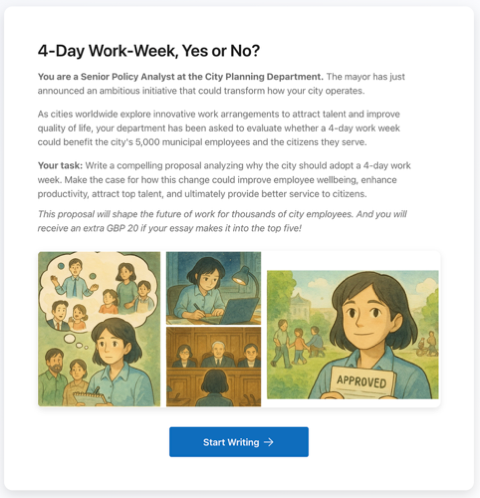}
\caption{Scenario brief 1 | \textbf{4-Day Work-Week, Yes or No?} Participants were cast as a senior policy analyst in a city-planning department and asked to write a proposal about whether a city should adopt a 4-day work week for municipal employees. The brief framed the task as a persuasive knowledge-work assignment requiring trade-off analysis, audience awareness and justification of recommendations.}
\label{fig:appendix_task1}
\end{figure}

\begin{figure}[htbp]
\centering
\includegraphics[width=0.86\linewidth]{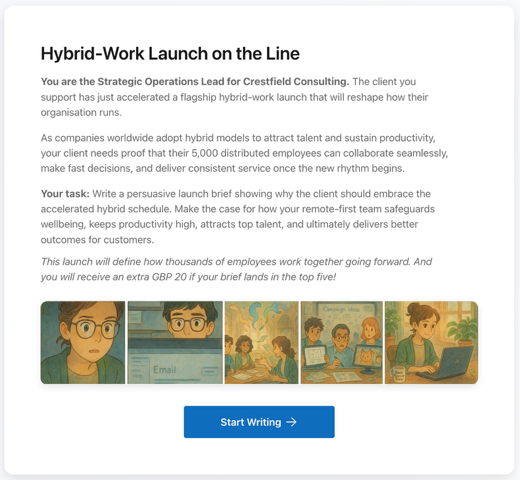}
\caption{Scenario brief 2 | \textbf{Hybrid-Work Launch on the Line.} Participants were cast as a strategic operations lead at a consulting firm and asked to write a persuasive brief for a client considering a major hybrid-work rollout. The brief emphasized decision quality, productivity, employee well-being, talent attraction and client outcomes.}
\label{fig:appendix_task2}
\end{figure}
\FloatBarrier

\FloatBarrier\clearpage
\section{Essay scoring: rubric and human-rating interface}\label{app:rubric}

\subsection{Essay-scoring rubric}
\autoref{tab:essay-scoring-rubric} shows the seven analytic criteria (A--G), each rated on a 1--4 scale, that were used by the human raters and supplied to the LLM judge. The human-rating interface is described in \autoref{app:rating-interface}.

\begin{table*}[htbp]
\centering
\caption{Essay-scoring rubric for the seven shared criteria (A--G), used by the human raters and supplied to the LLM judge.}
\label{tab:essay-scoring-rubric}
\scriptsize
\setlength{\tabcolsep}{4pt}
\renewcommand{\arraystretch}{1.15}
\begin{tabularx}{\textwidth}{@{}p{1.8cm}*{4}{>{\raggedright\arraybackslash}X}@{}}
\toprule
Criterion & Score 1 & Score 2 & Score 3 & Score 4 \\
\midrule
A. Context \& Purpose & Little/no problem framing; unclear audience; substantial off-task content. & Vague or generic framing; audience needs are mostly implicit; parts feel templated or only loosely tied to the decision context. & Frames the problem and audience reasonably well; mild mismatch in emphasis but remains on task. & Clearly frames the decision problem; explicitly tailors to the intended decision-maker; stays tightly aligned to the task. \\
\addlinespace
B. Recommendation Clarity & No clear recommendation/position. & Stance is implied or wobbly; proposal details are underspecified or contradictory. & Clear stance, but key specifics (scope/model) are incomplete, delayed, or mildly inconsistent. & States a clear recommendation early and specifies what is being proposed (scope, model, key design choices); provides a concise summary of the rationale. \\
\addlinespace
C. Argument Quality & Disorganized, internally inconsistent, or largely non-argumentative. & Relies on assertions or one-step reasoning; weak linkage between points. & Mostly logical and persuasive; some reasoning gaps or unexamined assumptions. & Builds a coherent chain from claims $\rightarrow$ reasons $\rightarrow$ support; makes assumptions/warrants legible. \\
\addlinespace
D. Evidence and Credibility & Virtually no support; repeated overclaims; credibility is questionable. & Mostly generalities; evidence is thin/generic; numbers (if any) feel ungrounded. & Some evidence, but uneven (claims without support in places). & Uses specific supporting material (examples, comparisons); avoids overclaims; support is relevant and credible. \\
\addlinespace
E. Feasibility & No implementation thinking; proposal would be difficult to execute as described. & Implementation is largely aspirational; feasibility concerns are waved away; limited operational detail. & Mentions implementation with some concrete steps; gaps remain (timeline vague, risks not fully addressed, success metrics thin). & Presents a realistic implementation path: key steps, timeline or pilot, operational coverage plan, governance/roles, and how success will be measured. \\
\addlinespace
F. Impact & Ignores stakeholder impacts or makes one-sided claims without considering harms. & Narrow focus (e.g., only employee wellbeing or only productivity); description of impacts are minimal. & Discusses stakeholder impacts, but not deeply; consequences lightly addressed. & Identifies key stakeholders (employees, managers); anticipates impacts and (unintended) consequences; links change to outcomes. \\
\addlinespace
G. Organization & Poorly organized; difficult to follow. & Some structure but hard to scan; repeated points; unclear prioritization. & Generally well organized; occasional digressions or weak transitions; still coherent and readable. & The sequence of ideas and semantic signposting make the recommendation, rationale and implementation easy to follow; information order supports decision-making and redundancy is minimal. \\
\bottomrule
\end{tabularx}
\par\smallskip
\end{table*}
\FloatBarrier

\clearpage

\subsection{Human-rating interface}\label{app:rating-interface}

\begin{figure}[htbp]
\centering
\includegraphics[width=\linewidth]{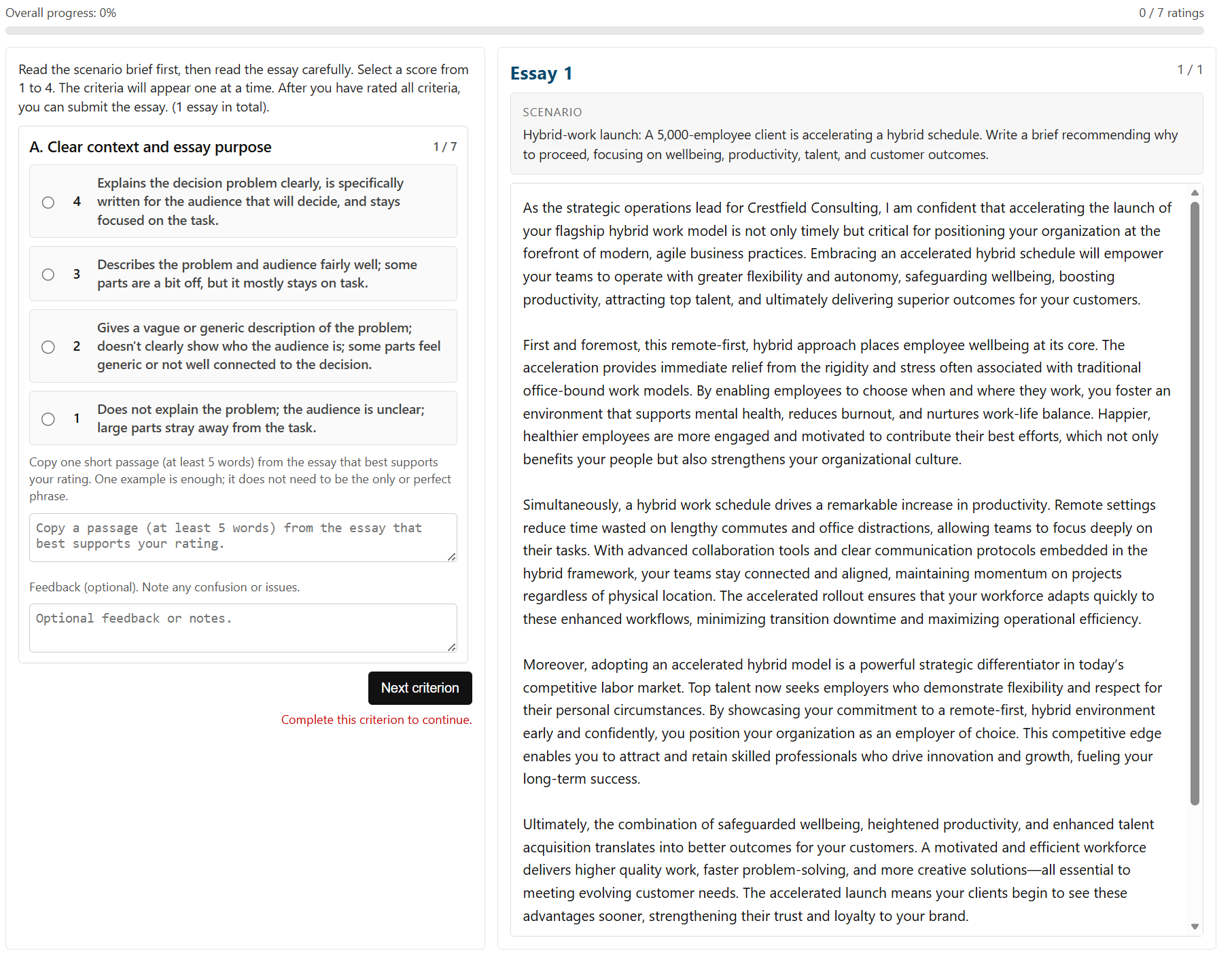}
\caption{Human-rating interface for scoring and citing supporting essay evidence.}
\label{fig:rating-interface}
\end{figure}

\FloatBarrier

\FloatBarrier\clearpage
\section{Prompt-coding codebook and coder agreement}\label{app:codebook}

\begin{table}[htbp]
\centering
\footnotesize
\caption{Prompt-coding reliability on the 60-sequence subset (original categorical codebook ratings), human--essay-judge agreement, and reliability across five essay-judge runs. R1--R3 are the three human coders; LLM is the automated coder applying the same codebook.}
\label{tab:rater-reliability}

\begin{tabular}{@{}llrrl@{}}
\toprule
Comparison & Measure & $N$ & Estimate & 95\% CI \\
\midrule
\multicolumn{5}{@{}l}{\textit{Prompt coding: goal clarity (0--3)}} \\
R1--R2 & $\kappa_w$ & 60 & 0.703 & [0.463, 0.844] \\
R1--R3 & $\kappa_w$ & 60 & 0.587 & [0.315, 0.755] \\
R2--R3 & $\kappa_w$ & 60 & 0.654 & [0.368, 0.813] \\
LLM--R1 & $\kappa_w$ & 60 & 0.458 & [0.269, 0.613] \\
LLM--R2 & $\kappa_w$ & 60 & 0.435 & [0.196, 0.604] \\
LLM--R3 & $\kappa_w$ & 60 & 0.740 & [0.510, 0.854] \\
LLM versus three-coder mean & ICC(A,1) & 60 & 0.621 & [0.403, 0.751] \\
Three-coder mean & ICC(A,3) & 60 & 0.849 & [0.686, 0.917] \\
\addlinespace
\multicolumn{5}{@{}l}{\textit{Prompt coding: task decomposition (0--3)}} \\
R1--R2 & $\kappa_w$ & 60 & 0.648 & [0.401, 0.831] \\
R1--R3 & $\kappa_w$ & 60 & 0.520 & [0.278, 0.697] \\
R2--R3 & $\kappa_w$ & 60 & 0.621 & [0.328, 0.818] \\
LLM--R1 & $\kappa_w$ & 60 & 0.363 & [0.059, 0.575] \\
LLM--R2 & $\kappa_w$ & 60 & 0.364 & [0.043, 0.606] \\
LLM--R3 & $\kappa_w$ & 60 & 0.482 & [0.172, 0.727] \\
LLM versus three-coder mean & ICC(A,1) & 60 & 0.517 & [0.155, 0.748] \\
Three-coder mean & ICC(A,3) & 60 & 0.817 & [0.647, 0.900] \\
\addlinespace
\multicolumn{5}{@{}l}{\textit{Prompt coding: combined GC+TD}} \\
R1--R2 & $\kappa_w$ & 60 & 0.677 & [0.447, 0.837] \\
R1--R3 & $\kappa_w$ & 60 & 0.578 & [0.376, 0.723] \\
R2--R3 & $\kappa_w$ & 60 & 0.699 & [0.479, 0.830] \\
LLM--R1 & $\kappa_w$ & 60 & 0.462 & [0.287, 0.602] \\
LLM--R2 & $\kappa_w$ & 60 & 0.517 & [0.297, 0.663] \\
LLM--R3 & $\kappa_w$ & 60 & 0.681 & [0.488, 0.799] \\
LLM versus three-coder mean & ICC(A,1) & 60 & 0.662 & [0.482, 0.780] \\
Three-coder mean & ICC(A,3) & 60 & 0.848 & [0.714, 0.915] \\
\midrule
\multicolumn{5}{@{}l}{\textit{Human--essay-judge agreement: seven shared dimensions}} \\
Human versus judge & ICC(A,1) & 107 & 0.456 & [0.349, 0.547] \\
\midrule
\multicolumn{5}{@{}l}{\textit{Essay-judge inter-run reliability: seven-dimension composite}} \\
Single run & ICC(A,1) & 594 & 0.711 & [0.666, 0.747] \\
Mean of five runs & ICC(A,5) & 594 & 0.925 & [0.909, 0.937] \\
\bottomrule
\end{tabular}
\end{table}

\subsection{Prompt-coding codebook}

\textbf{Purpose of the codebook.} This codebook was used to structure the manual prompt-coding stage that preceded LLM-based extension of the prompt annotations. It was designed to capture metacognitive properties of participant prompting, especially the extent to which conversations made the goal of the writing task explicit and decomposed the task into manageable sub-goals. The framework draws on Flower and Hayes' cognitive process account of writing, hierarchical task analysis, cognitive task analysis, and the 4D AI Fluency framework.

\textbf{Coding workflow.} The unit of analysis was the participant-authored prompt sequence within an AI-assisted writing session. A 20\% subset of sequences (60 conversations) was independently coded by three researchers using the codebook below, and an LLM coder applied the same codebook to the same sequences; these codes were used to assess inter-coder reliability and the LLM coder's agreement with the human coders (\autoref{tab:rater-reliability}). The graded indicator scoring used in the analyses is described in Methods.

\subsection{A. Goal clarity}

\textbf{Construct.} Goal clarity captures how explicitly the conversation communicates a clear purpose or intended outcome for the writing task. Coders assess the full prompt sequence holistically to determine how clearly the user specifies what the writing should achieve, for whom, and under what constraints.

\begin{table}[htbp]
\centering
\caption{Goal-clarity dimension-level codes}
\label{tab:gc-codes}
\small
\begin{tabularx}{\linewidth}{@{}L{1.1cm}L{2.2cm}Y Y@{}}
\toprule
Code & Label & Definition & Coding cue \\
\midrule
GC-0 & Absent & No discernible goal or purpose is expressed. Prompts consist of bare topics or instructions so vague that the intended writing outcome cannot be inferred. & No explicit purpose, audience or constraint. \\
GC-1 & Implicit & A general purpose can be inferred from context, but it is never stated explicitly. The user delegates full goal definition to the AI. & Genre or topic implies a goal, but the user does not specify it. \\
GC-2 & Moderate & At least one explicit goal is stated, but the conversation does not specify audience, constraints or evaluative criteria in enough detail to make the target unambiguous. & Purpose or genre is named, but supporting constraints are thin. \\
GC-3 & High & Clear purpose is articulated plus at least two of: audience, constraints, format/tone or evaluative criteria. The intended outcome is unambiguous and may develop across turns. & A full rhetorical problem is made explicit. \\
\bottomrule
\end{tabularx}
\end{table}

\textbf{Goal-clarity indicators.} For conversations coded GC-1 or above, coders optionally marked whether the following indicators were present anywhere in the full prompt sequence: purpose stated (GC-P), audience specified (GC-A), constraints articulated (GC-C), evaluative criteria (GC-E), genre/form named (GC-G), rhetorical situation (GC-R), and process goal (GC-PG).

\textbf{Goal-clarity decision rules.} GC-0 was used when no indicators were present. GC-1 was used when the goal was only implicitly derivable. GC-2 required at least one explicit indicator (typically purpose or genre). GC-3 required an explicit purpose plus at least two additional indicators such as audience, constraints or evaluative criteria.

\subsection{B. Task decomposition}

\textbf{Construct.} Task decomposition captures the extent to which the prompt sequence breaks the writing task into identifiable sub-tasks, steps or structured parts rather than presenting it as a single undifferentiated request.

\begin{table}[htbp]
\centering
\caption{Task-decomposition dimension-level codes}
\label{tab:td-codes}
\small
\begin{tabularx}{\linewidth}{@{}L{1.1cm}L{2.3cm}Y Y@{}}
\toprule
Code & Label & Definition & Coding cue \\
\midrule
TD-0 & Monolithic & The writing task is presented as a single undifferentiated request. No sub-tasks, steps or structural divisions are discernible. & ``Write about climate change.'' \\
TD-1 & Partially decomposed & More than one component or aspect of the task is named or implied, but without specified ordering or relations among those components. & Components are listed but not organized. \\
TD-2 & Structured & The task is explicitly organized into ordered steps, labeled sections or a sequence of sub-tasks. Relationships between parts are specified at one level. & Numbered or ordered steps are present. \\
TD-3 & Hierarchical & The task is decomposed at multiple levels: top-level goals are broken into sub-goals, which are themselves broken into operations or priorities. Dependencies across levels are indicated. & Multi-level structure with priorities or dependencies. \\
\bottomrule
\end{tabularx}
\end{table}

\textbf{Task-decomposition indicators.} For conversations coded TD-1 or above, coders optionally marked whether the prompt sequence identified sub-goals (TD-SG), specified sequencing (TD-SQ), indicated differential granularity (TD-GR), noted dependencies among parts (TD-DP), specified delegation structure between user and AI (TD-DL), or displayed iterative structure across turns (TD-IT).

\textbf{Task-decomposition decision rules.} TD-0 was used when no decomposition was visible. TD-1 required identifiable sub-goals without an explicit sequence. TD-2 required both sub-goals and sequencing. TD-3 required sub-goals and sequencing plus at least two additional indicators such as granularity, dependencies, delegation structure or iteration.

\subsection{C. Worked examples}

\textbf{Example 1: minimal single-turn prompt.} ``Write about climate change.'' This was coded GC-0 and TD-0 because neither the goal nor the structure of the task was made explicit.

\textbf{Example 2: goal-rich but undecomposed prompt.} ``Write a persuasive blog post about remote work for HR managers. Keep it under 800 words.'' This was coded GC-3 because it specifies purpose, audience, genre and a length constraint, but TD-0 because the task is still presented monolithically.

\textbf{Example 3: iterative decomposed prompt sequence.} ``I need a policy brief for the Minister of Education on digital literacy funding. Start with an executive summary, then three evidence-based arguments with statistics, then address the main counterargument, and end with a policy recommendation. Formal but accessible tone. ... Now strengthen the second argument with OECD data. ... I will handle the final edit; tighten the executive summary to 100 words.'' This was coded GC-3 and TD-3 because the goal is rhetorically explicit and the task is decomposed hierarchically across multiple turns.

\subsection{D. Coding procedure and edge cases}

The coding procedure proceeded in six steps: (1) read the full prompt sequence; (2) assign a goal-clarity rating; (3) mark goal-clarity indicators; (4) assign a task-decomposition rating; (5) mark task-decomposition indicators; and (6) record coder notes for edge cases.

Because the coding was conversation-level, indicators could appear in any turn. A goal stated in an early turn and a constraint added later both counted. Decomposition that emerged progressively across turns was coded at the highest level observed in the full sequence.

Three edge-case rules were applied. First, if the goal had to be inferred only from genre conventions, coding was capped at GC-1. Second, bullet points used only for readability did not count as task decomposition unless they clearly structured sub-tasks. Third, AI-proposed structure did not count as user-initiated decomposition unless the user modified, elaborated or otherwise actively took it up.

\subsection{E. Blank coding sheet}

\begin{table}[htbp]
\centering
\caption{Blank coding sheet}
\label{tab:coding-sheet}
\small
\begin{tabularx}{\linewidth}{@{}L{4.4cm}Y@{}}
\toprule
Field & Value \\
\midrule
Conversation ID & \\
Number of user turns & \\
Writing task type & \\
GC rating & GC-0 / GC-1 / GC-2 / GC-3 \\
GC-P (purpose) & 0 / 1 \\
GC-A (audience) & 0 / 1 \\
GC-C (constraints) & 0 / 1 \\
GC-E (evaluative criteria) & 0 / 1 \\
GC-G (genre/form) & 0 / 1 \\
GC-R (rhetorical situation) & 0 / 1 \\
GC-PG (process goal) & 0 / 1 \\
TD rating & TD-0 / TD-1 / TD-2 / TD-3 \\
TD-SG (sub-goals) & 0 / 1 \\
TD-SQ (sequencing) & 0 / 1 \\
TD-GR (granularity) & 0 / 1 \\
TD-DP (dependencies) & 0 / 1 \\
TD-DL (delegation) & 0 / 1 \\
TD-IT (iterative structure) & 0 / 1 \\
Coder notes & \\
\bottomrule
\end{tabularx}
\end{table}

\FloatBarrier\clearpage

\section{Survey items}\label{app:survey-items}

All items administered in the four surveys are listed in \autoref{tab:survey-items} in the order in which participants encountered them: the baseline survey before Task~1, the post-task survey after Task~1, the debrief survey after Task~2, and the voluntary follow-up survey three days later.

\begingroup
\footnotesize
\setlength{\tabcolsep}{4pt}
\renewcommand{\arraystretch}{1.1}
\begin{longtable}{@{}>{\raggedright\arraybackslash}p{0.10\linewidth}>{\raggedright\arraybackslash}p{0.86\linewidth}@{}}
\caption{Survey items by stage, in the order administered. Response scales are given in the block headings or in square brackets after the item. Items shown with a leading ellipsis were presented under a shared introductory stem in the survey form. Item numbers within blocks match those used in \autoref{tab:metacognition-items} and \autoref{tab:missingness-firth}. Participant-identifier fields and the completion-code screen are omitted.}\label{tab:survey-items}\\
\toprule
Item & Wording [response options] \\
\midrule
\endfirsthead
\multicolumn{2}{@{}l}{\textit{Survey items (continued)}}\\
\toprule
Item & Wording [response options] \\
\midrule
\endhead
\midrule
\endfoot
\bottomrule
\endlastfoot
\addlinespace[5pt]
\multicolumn{2}{@{}l}{\textbf{Baseline survey (before Task~1)}} \\
\addlinespace[3pt]
\multicolumn{2}{@{}p{0.97\linewidth}@{}}{\textit{Prior AI use}} \\
AI-use~1 & How often do you use AI tools like ChatGPT? [once per month / once per week / a few times per week / daily / a few times daily] \\
AI-use~2 & What do you use AI tools like ChatGPT for? [professional writing / creative writing / research / coding / general knowledge / content creation / other (free text)] \\
\addlinespace[3pt]
\multicolumn{2}{@{}p{0.97\linewidth}@{}}{\textit{General metacognitive regulation, adapted from the Metacognitive Awareness Inventory~\cite{SchrawDennison1994} (1 = does not describe me at all, 5 = describes me very much)}} \\
Meta~1 & I set specific goals before I begin a task \\
Meta~2 & I ask myself periodically if I am meeting my goals. \\
Meta~3 & I organise my time to best accomplish my goals. \\
Meta~4 & I ask myself how well I accomplished my goals once I am finished \\
Meta~5 & I try to break tasks down into smaller steps. \\
Meta~6 & I ask myself questions about the material before I begin. \\
Meta~7 & I think about what I really need to do before I begin a task. \\
Meta~8 & I read instructions carefully before I begin a task. \\
\addlinespace[3pt]
\multicolumn{2}{@{}p{0.97\linewidth}@{}}{\textit{Writing metacognition, adapted from the Metacognitive Strategies in Writing questionnaire~\cite{QinZhangXiao2022} and the Metacognitive Strategies of Writing Survey~\cite{AlMoqbali2020} (1 = strongly disagree, 5 = strongly agree)}} \\
Meta~9 & \ldots I check if my writing meets the requirements of the writing task. \\
Meta~10 & \ldots I check if the content of my writing is relevant to the topic. \\
Meta~11 & \ldots I check if the organization of my writing is logical. \\
Meta~12 & \ldots I check if the language of my writing is clear enough. \\
Meta~13 & \ldots my composing process is usually like this: Analysing the writing task, generating ideas, planning, outlining, drafting, and revising. \\
\addlinespace[3pt]
\multicolumn{2}{@{}p{0.97\linewidth}@{}}{\textit{Self-rated writing skill}} \\
Skill & How would you rate your essay writing skills? [beginner / novice / intermediate / advanced / expert] \\
\addlinespace[3pt]
\multicolumn{2}{@{}p{0.97\linewidth}@{}}{\textit{AI metacognition, items from \citet{sidra_generative_2025} (1 = never, 7 = always)}} \\
AI-meta~1 & \ldots I think about how to give the most appropriate information to the AI. \\
AI-meta~2 & \ldots I consider what information the AI needs to perform the task. \\
AI-meta~3 & \ldots I think about how well I understand what the AI is good and not so good at for the task we're doing. \\
AI-meta~4 & \ldots I think about how best to divide the work between me and the AI. \\
AI-meta~5 & \ldots I review my thinking to ensure that I'm not accepting or rejecting the AI's recommendation too quickly. \\
AI-meta~6 & \ldots I check task goals every now and then to make sure we're on track. \\
AI-meta~7 & \ldots I try to ensure that my own mental shortcuts or biases aren't affecting my reactions to the AI's output. \\
AI-meta~8 & \ldots I remind myself about my own limitations or weaknesses so as to minimise their impact on the task. \\
AI-meta~9 & \ldots I think about how I can improve the way I work with AI. \\
AI-meta~10 & \ldots I analyse my thinking before jumping to conclusions regarding the AI's output. \\
AI-meta~11 & \ldots If something goes wrong, I analyse the situation so I can avoid a similar error in the future. \\
\addlinespace[3pt]
\multicolumn{2}{@{}p{0.97\linewidth}@{}}{\textit{Prompting self-efficacy, items from \citet{gibreel_development_2025} (1 = strongly disagree, 5 = strongly agree)}} \\
Prompt~1 & \ldots I can break down complex tasks into well-structured prompts, guiding AI step by step while minimizing trial and error. \\
Prompt~2 & \ldots I can adjust my prompts using different formats (e.g., structured lists, open-ended questions, scenario-based instructions) to enhance response quality and relevance. \\
Prompt~3 & \ldots I can improve AI-generated responses by specifying constraints such as role, tone, length, format, or other contextual factors. \\
Prompt~4 & \ldots I can improve response relevance by incorporating real world examples, hypothetical scenarios, or case studies tailored to different AI tools. \\
\addlinespace[5pt]
\multicolumn{2}{@{}l}{\textbf{Post-task survey (after Task~1)}} \\
\addlinespace[3pt]
\multicolumn{2}{@{}p{0.97\linewidth}@{}}{\textit{Task metacognition, post-task wording of the Metacognitive Awareness Inventory items (1 = does not describe me at all, 5 = describes me very much)}} \\
Meta~1 & I set specific goals before I began with the task \\
Meta~2 & I asked myself periodically if I was meeting my goals. \\
Meta~3 & I organised my time to best accomplish my goals. \\
Meta~4 & I asked myself how well I accomplished my goals once I was finished \\
Meta~5 & I tried to break tasks down into smaller steps. \\
Meta~6 & I asked myself questions about the task before I began. \\
Meta~7 & I thought about what I really needed to do before I began the task. \\
Meta~8 & I read instructions carefully before I began the task. \\
\addlinespace[3pt]
\multicolumn{2}{@{}p{0.97\linewidth}@{}}{\textit{Writing metacognition, post-task wording of the writing-metacognition items (1 = strongly disagree, 5 = strongly agree)}} \\
Meta~9 & \ldots I checked if my essay met the requirements of the writing task. \\
Meta~10 & \ldots I checked if the content of my essay was relevant to the topic. \\
Meta~11 & \ldots I checked if the organization of my essay was logical. \\
Meta~12 & \ldots I checked if the language of my essay was clear enough. \\
Meta~13 & \ldots my composing process was like this: I analysed the writing task, then I generated ideas, then I planned the essay, then I outlined the essay, drafter, and revised it. \\
\addlinespace[3pt]
\multicolumn{2}{@{}p{0.97\linewidth}@{}}{\textit{AI experience}} \\
Help & How helpful did you find the AI tool? [1--5] \\
Control & How much in control did you feel when using the AI tool? [1--5] \\
Open & What aspects of the AI tool were most beneficial? Which were least beneficial? [open-ended] \\
\addlinespace[3pt]
\multicolumn{2}{@{}p{0.97\linewidth}@{}}{\textit{Workload: NASA-TLX items (1--10)}} \\
Load~1 & How mentally demanding was the essay writing task? \\
Load~2 & How hurried or rushed was the pace of the task? \\
Load~3 & How successful were you in accomplishing what you were asked to do? \\
Load~4 & How hard did you have to work to accomplish your level of performance? \\
Load~5 & How insecure, discouraged, irritated, stressed, or annoyed were you? \\
\addlinespace[3pt]
\multicolumn{2}{@{}p{0.97\linewidth}@{}}{\textit{Workload: additional AI-writing demand items (1--10)}} \\
Load~6 & How mentally demanding was the essay topic itself to understand and write about? \\
Load~7 & How much mental effort did developing coherent arguments for this essay demand? \\
Load~8 & How much mental burden did managing the AI tool add for writing the essay? \\
Load~9 & How mentally demanding was prompting the AI? \\
Load~10 & Did you have to think hard about how to structure your essay effectively? \\
Load~11 & Did integrating AI suggestions require careful mental processing? \\
Load~12 & Did you struggle to keep track of all the ideas you wanted to include? \\
Load~13 & Was it mentally challenging to maintain consistency throughout the essay? \\
Load~14 & Was it difficult to maintain focus throughout the task? \\
\addlinespace[5pt]
\multicolumn{2}{@{}l}{\textbf{Debrief survey (after Task~2)}} \\
\addlinespace[3pt]
\multicolumn{2}{@{}p{0.97\linewidth}@{}}{\textit{Workload (1--10)}} \\
Load~1--14 & Same 14 workload items as after Task~1, referring to the second essay; wording identical except Load~1 (``How mentally demanding was the essay-writing task?'') and Load~8 (``How much mental burden did managing the AI tool while writing add?''). \\
\addlinespace[3pt]
\multicolumn{2}{@{}p{0.97\linewidth}@{}}{\textit{Workflow and future AI use}} \\
Workflow & Tell us about your workflow with the AI tool [open-ended] \\
Future~use & If you had to complete a similar task in the future, would you use the AI tool? (If yes, why? If no, why not?) [open-ended] \\
\addlinespace[3pt]
\multicolumn{2}{@{}p{0.97\linewidth}@{}}{\textit{Follow-up invitation}} \\
Invite~1 & To confirm, are you aware that completing the follow-up survey in 3 days' time (on Thursday) will make you eligible to win one of 20 extra rewards on Prolific over \pounds30 (or local equivalent, e.g.\ approx.\ \$40)? [yes / no; remainder of the question text abridged] \\
Invite~2 & Are you able and willing to complete the follow-up study in 3 days (on Thursday)? [yes / no] \\
\addlinespace[3pt]
\multicolumn{2}{@{}p{0.97\linewidth}@{}}{\textit{Closing}} \\
Comments & Closing message thanking participants and inviting any additional comments [open-ended] \\
\addlinespace[5pt]
\multicolumn{2}{@{}l}{\textbf{Three-day follow-up survey}} \\
\addlinespace[3pt]
\multicolumn{2}{@{}p{0.97\linewidth}@{}}{\textit{Topic recall}} \\
Recall~1 & What topic did you write about in your first essay? If you don't remember, you can write ``Don't remember''. [open-ended] \\
Recall~2 & What topic did you write about in your second essay? If you don't remember, you can write ``Don't remember''. [open-ended] \\
\addlinespace[3pt]
\multicolumn{2}{@{}p{0.97\linewidth}@{}}{\textit{Free recall of proposal content}} \\
Recall~3 & Write down everything you remember about your proposal for a 4-day work week -- for example the main sections you covered, which information you included, or key topics that you wrote about. If you don't remember, you can write ``Don't remember''. [open-ended] \\
Recall~4 & Write down everything you remember about your proposal for remote work -- for example the main sections you covered, which information you included, or key topics that you wrote about. If you don't remember, you can write ``Don't remember''. [open-ended] \\
\addlinespace[3pt]
\multicolumn{2}{@{}p{0.97\linewidth}@{}}{\textit{Recognition checklists (select all that apply)}} \\
Recog~1 & Which benefits of a 4-day work week did you discuss? Select all that apply and/or specify ``other'' (note: not all will apply) [attractive for top talent / better productivity / less burn-out / improved customer satisfaction / rising revenue / employee retention / fewer meetings / long-term success / improved well-being / reduced office cost / higher job turnover / none of the above / I don't remember / other (please specify)] \\
Recog~2 & Which benefits of a remote-work model did you discuss? Select all that apply and/or specify ``other'' (note: not all will apply) [stronger team cohesion / reduced stress level / better productivity / fewer sick days / time saving / better communication / money saving / environmental benefits / talent access / improved work-life balance / none of the above / I don't remember / other (please specify)] \\
\addlinespace[3pt]
\multicolumn{2}{@{}p{0.97\linewidth}@{}}{\textit{Reflection}} \\
Insight & What is one insight about these topics that surprised you most? [open-ended] \\
\end{longtable}
\endgroup

\FloatBarrier\clearpage

\section{Supplementary results: essay quality}\label{app:res-quality}

\begin{table}[htbp]
\centering
\caption{Mixed-effects models for primary and time-on-task robustness models, adjusted for the six common baseline covariates.}
\label{tab:essay-models}
\footnotesize
\setlength{\tabcolsep}{4pt}
\begin{tabular}{@{}p{4.6cm}rrrrrr@{}}
\toprule
& \multicolumn{3}{c}{Primary model}
& \multicolumn{3}{c}{+ log time on task} \\
\cmidrule(lr){2-4}\cmidrule(lr){5-7}
Term or contrast & Coef. & s.e. & $P$ & Coef. & s.e. & $P$ \\
\midrule
Intercept & $2.753$ & 0.088 & $<.001$ & $2.761$ & 0.157 & $<.001$ \\
Task~1: scaffold vs control & $0.089$ & 0.034 & .008 & $0.089$ & 0.035 & .011 \\
Scaffold $\times$ second task & $-0.019$ & 0.043 & .665 & $-0.019$ & 0.044 & .663 \\
Second vs first task (control) & $0.024$ & 0.031 & .440 & $0.023$ & 0.032 & .472 \\
Topic: city workweek vs hybrid launch & $0.251$ & 0.021 & $<.001$ & $0.251$ & 0.022 & $<.001$ \\
Baseline writing purpose & $-0.024$ & 0.034 & .478 & $-0.024$ & 0.034 & .477 \\
Baseline writing skill & $0.003$ & 0.015 & .829 & $0.003$ & 0.015 & .826 \\
Baseline task decomposition & $-0.002$ & 0.014 & .912 & $-0.002$ & 0.014 & .914 \\
Baseline AI evaluating & $-0.003$ & 0.010 & .783 & $-0.003$ & 0.010 & .783 \\
Baseline AI prompting & $0.007$ & 0.011 & .554 & $0.007$ & 0.011 & .553 \\
Baseline AI-use frequency & $-0.0004$ & 0.011 & .974 & $-0.0004$ & 0.011 & .975 \\
Time on task (log seconds) & --- & --- & --- & $-0.001$ & 0.021 & .951 \\
\midrule
Task~2: previously scaffolded vs previously unscaffolded & $0.070$ & 0.034 & .037 & $0.070$ & 0.034 & .037 \\
\midrule
Participant random-intercept variance & 0.0146 & --- & --- & 0.0147 & --- & --- \\
Residual variance & 0.0679 & --- & --- & 0.0680 & --- & --- \\
Essays & 594 & --- & --- & 594 & --- & --- \\
Participants & 297 & --- & --- & 297 & --- & --- \\
\bottomrule
\end{tabular}
\end{table}

\begin{table}[htbp]
\centering
\caption{Agreement between the LLM essay judge and the human benchmark on the 107 human-rated essays not used as calibration examples. Human scores are the mean of two raters on the formatted essays; judge scores are the mean of five runs on the de-formatted essays. All statistics are computed on essay-level means across the seven shared rubric dimensions (A--G). The calibration slope is from a regression of judge on human scores; a slope below 1 indicates that the judge compresses the range of human scores.}
\label{tab:judge-agreement}
\small
\begin{tabular}{@{}L{6.6cm}C{1.6cm}C{2.4cm}@{}}
\toprule
Metric & Estimate & 95\% CI \\
\midrule
Essays compared & 107 & --- \\
Absolute-agreement ICC(A,1), human versus judge & 0.456 & [0.349, 0.547] \\
Pearson $r$, human versus judge & 0.662 & --- \\
Calibration slope, judge on human & 0.375 & --- \\
\bottomrule
\end{tabular}
\end{table}

\begin{table*}[htbp]
\centering
\footnotesize
\caption{Standardized essay-quality effect sizes (de-formatted judge; $n=297$)}
\label{tab:standardized-effects}
\setlength{\tabcolsep}{3pt}
\begin{tabular}{@{}lrrrrrr@{}}
\toprule
Task & \shortstack{Scaffold-assigned\\$M$ (SD)}
& \shortstack{Control\\$M$ (SD)}
& \shortstack{Raw\\difference}
& Hedges' $g$
& \shortstack{Adjusted\\difference}
& \shortstack{Adjusted standardized\\difference} \\
\midrule
Task 1 & 2.99 (0.32) & 2.90 (0.32) & 0.094 & 0.29 & 0.089 & 0.28 \\
Task 2 & 2.98 (0.29) & 2.92 (0.32) & 0.063 & 0.21 & 0.070 & 0.23 \\
\bottomrule
\end{tabular}
\end{table*}

\FloatBarrier\clearpage

\section{Supplementary results: topic recall and follow-up missingness}\label{app:res-recall}

\begin{table}[htbp]
\centering
\caption{Topic-recall model summaries. OR is the treatment effect on the odds of correct topic recall (scaffold versus control for Task~1; previously scaffolded versus control for Task~2), from logistic mixed models with a participant random intercept, condition, task position, their interaction, scenario and the six common covariates (\autoref{subsubsec:covariates}). Observed rates are control versus scaffold.}
\label{tab:topic-recall-models}
\small
\setlength{\tabcolsep}{5pt}
\begin{tabular}{@{}L{2.0cm}L{1.2cm}L{3.0cm}L{2.6cm}C{1.1cm}L{2.4cm}@{}}
\toprule
Outcome & Task & Sample & OR [95\% CI] & $P$ & Observed rates (control vs scaffold) \\
\midrule
Topic recall & Task~1 & Complete case & $5.07$ [$1.21$, $21.25$] & .026 & 45.6\% vs 64.6\% \\
Topic recall & Task~1 & Multiple imputation ($m=50$) & $3.24$ [$1.11$, $9.49$] & .032 & --- \\
Topic recall & Task~2 & Complete case & $4.24$ [$1.02$, $17.57$] & .046 & 38.0\% vs 54.4\% \\
Topic recall & Task~2 & Multiple imputation ($m=50$) & $3.06$ [$1.04$, $9.02$] & .042 & --- \\
\bottomrule
\end{tabular}
\par\smallskip
\end{table}

\begingroup
\footnotesize
\setlength{\tabcolsep}{3pt}
\renewcommand{\arraystretch}{1.12}
\begin{longtable}{@{}>{\raggedright\arraybackslash}p{0.23\linewidth}>{\raggedright\arraybackslash}p{0.36\linewidth}>{\centering\arraybackslash}p{0.07\linewidth}>{\centering\arraybackslash}p{0.17\linewidth}>{\centering\arraybackslash}p{0.07\linewidth}@{}}
\caption{Firth-penalized logistic model of missing scored recall records: predictors, scoring and estimates. OR is the change in the odds of a participant's recall record being missing, per SD of each standardized continuous predictor or for the coded category of each binary predictor, with all predictors entered jointly ($N=234$ participants with complete predictor data); OR $>1$ indicates higher odds of missingness. $\alpha$: Cronbach's alpha of composite predictors; NE: not estimated. Six further candidate indicators (writing-cohort membership, ambiguous exposure, missing post-task form, missing debrief form, Task~1 and Task~2 essay submitted) were constant across participants and were excluded.}\label{tab:missingness-firth}\\
\toprule
Predictor & Definition / scoring & $\alpha$ & OR [95\% CI] & $P$ \\
\midrule
\endfirsthead
\multicolumn{5}{l}{\textit{Firth-penalized model of missing scored recall records (continued)}}\\
\toprule
Predictor & Definition / scoring & $\alpha$ & OR [95\% CI] & $P$ \\
\midrule
\endhead
\midrule
\endfoot
\bottomrule
\endlastfoot
Treatment & Scaffold = 1; control = 0 & -- & $1.65$ [$0.75$, $3.73$] & .216 \\
First-task scenario & City workweek first = 1; hybrid-work first = 0 & -- & $1.55$ [$0.73$, $3.32$] & .252 \\
AI-use frequency & Single baseline item, 1--5 & -- & $1.18$ [$0.88$, $1.59$] & .264 \\
AI use for writing & Writing-related AI use = 1; otherwise = 0 & -- & $1.01$ [$0.49$, $2.05$] & .983 \\
Essay-writing skill & Single baseline item, 1--5 & -- & $1.02$ [$0.76$, $1.36$] & .895 \\
Baseline general metacognition & Mean of metacognition items 1--8; 1--5 & 0.862 & $1.27$ [$0.88$, $1.85$] & .208 \\
Baseline writing metacognition & Mean of metacognition items 9--13; 1--5 & 0.758 & $1.03$ [$0.70$, $1.52$] & .877 \\
Baseline AI metacognition & Mean of all 11 AI-metacognition items; 1--7 & 0.862 & $1.06$ [$0.77$, $1.47$] & .712 \\
Baseline prompting self-report & Mean of all four prompting items; 1--5 & 0.818 & $1.12$ [$0.81$, $1.57$] & .493 \\
Post-task general metacognition & Mean of Task 1 post-task items 1--8; 1--5 & 0.882 & $0.98$ [$0.63$, $1.51$] & .922 \\
Post-task writing metacognition & Mean of Task 1 post-task items 9--13; 1--5 & 0.756 & $0.73$ [$0.49$, $1.08$] & .114 \\
Task 1 helpfulness & Single item, 1--5 & -- & $1.16$ [$0.80$, $1.69$] & .434 \\
Task 1 perceived control & Single item, 1--5 & -- & $0.95$ [$0.65$, $1.39$] & .790 \\
Task 1 NASA-TLX & Mean of load items 1--5; performance reversed; 1--10 & 0.690 & $1.03$ [$0.60$, $1.80$] & .905 \\
Task 2 NASA-TLX & Mean of load items 1--5; performance reversed; 1--10 & 0.758 & $1.02$ [$0.63$, $1.66$] & .927 \\
Task 1 AI-writing demands & Mean of supplementary load items 6--14; 1--10 & 0.913 & $0.87$ [$0.49$, $1.53$] & .632 \\
Task 2 AI-writing demands & Mean of supplementary load items 6--14; 1--10 & 0.924 & $0.95$ [$0.54$, $1.66$] & .850 \\
Task 1 essay quality & Mean across rubric criteria A--G and five judge runs; 1--4 & NE & $0.67$ [$0.47$, $0.94$] & .021 \\
Task 2 essay quality & Mean across rubric criteria A--G and five judge runs; 1--4 & NE & $0.90$ [$0.66$, $1.23$] & .511 \\
Task 2 combined prompt quality & GC + TD: sum of 13 indicator means; 0--39 & NE & $1.36$ [$1.02$, $1.80$] & .033 \\
Task 1 duration & Natural log of duration in seconds & -- & $0.91$ [$0.67$, $1.22$] & .504 \\
Task 2 duration & Natural log of duration in seconds & -- & $1.16$ [$0.85$, $1.59$] & .352 \\
Task 1 prompt count & Natural log of 1 + prompt count & -- & $1.11$ [$0.75$, $1.64$] & .609 \\
Task 2 prompt count & Natural log of 1 + prompt count & -- & $1.06$ [$0.76$, $1.49$] & .712 \\
Task 1 essay length & Natural log of 1 + word count & -- & $0.93$ [$0.66$, $1.31$] & .673 \\
Task 2 essay length & Natural log of 1 + word count & -- & $0.98$ [$0.71$, $1.37$] & .920 \\
\end{longtable}
\endgroup

\FloatBarrier\clearpage

\section{Supplementary results: subjective outcomes}\label{app:res-subjective}

\begin{table*}
\centering
\footnotesize
\caption{Item-level self-reported metacognition before and after Task~1}
\label{tab:metacognition-items}
\resizebox{\textwidth}{!}{%
\begin{tabular}{@{}cllccccrrrrr@{}}
\toprule
No. & Source & Item & Control pre & Control post & Scaffold pre & Scaffold post & $b$ & 95\% CI & $P$ & $P_{\mathrm{BH}}$ & $N$ \\
\midrule
1 & MAI & Set specific goals & 3.72 (0.89) & 3.77 (1.10) & 3.68 (0.95) & 3.91 (1.00) & $0.190$ & [-0.019, 0.398] & .074 & .241 & 290 \\
2 & MAI & Monitor goal progress & 3.68 (0.99) & 3.91 (1.01) & 3.59 (1.06) & 3.81 (0.96) & $-0.051$ & [-0.257, 0.154] & .624 & .811 & 280 \\
3 & MAI & Organize time around goals & 3.86 (0.94) & 3.83 (0.97) & 3.79 (1.04) & 3.68 (1.09) & $-0.090$ & [-0.311, 0.132] & .427 & .710 & 280 \\
4 & MAI & Evaluate goal attainment & 3.88 (0.91) & 4.20 (0.80) & 3.73 (0.99) & 4.03 (0.82) & $-0.102$ & [-0.265, 0.061] & .220 & .476 & 279 \\
5 & MAI & Break task into steps & 3.83 (1.00) & 3.24 (1.16) & 3.98 (0.90) & 4.03 (0.84) & $0.753$ & [0.546, 0.959] & $<.001$ & $<.001$ & 295 \\
6 & MAI & Ask questions before starting & 3.71 (1.06) & 3.89 (1.09) & 3.70 (1.00) & 3.87 (0.97) & $0.024$ & [-0.182, 0.229] & .821 & .931 & 292 \\
7 & MAI & Identify task requirements & 4.03 (0.87) & 4.08 (0.96) & 4.07 (0.89) & 4.13 (0.92) & $0.067$ & [-0.133, 0.267] & .511 & .739 & 291 \\
8 & MAI & Read instructions carefully & 4.33 (0.89) & 4.48 (0.80) & 4.36 (0.82) & 4.47 (0.74) & $-0.013$ & [-0.158, 0.132] & .859 & .931 & 292 \\
9 & MSW/MSWS & Check task requirements & 4.46 (0.67) & 4.58 (0.69) & 4.52 (0.60) & 4.45 (0.69) & $-0.148$ & [-0.299, 0.002] & .054 & .233 & 292 \\
10 & MSW/MSWS & Check content relevance & 4.63 (0.60) & 4.65 (0.60) & 4.62 (0.51) & 4.56 (0.64) & $-0.085$ & [-0.220, 0.050] & .216 & .476 & 291 \\
11 & MSW/MSWS & Check logical organization & 4.59 (0.56) & 4.55 (0.66) & 4.32 (0.79) & 4.47 (0.66) & $-0.062$ & [-0.217, 0.094] & .437 & .710 & 290 \\
12 & MSW/MSWS & Check language clarity & 4.59 (0.56) & 4.64 (0.56) & 4.44 (0.73) & 4.60 (0.59) & $-0.001$ & [-0.130, 0.128] & .987 & .987 & 292 \\
13 & MSW & Follow staged composing process & 4.13 (0.82) & 3.91 (1.17) & 3.99 (0.97) & 4.07 (0.90) & $0.225$ & [0.010, 0.440] & .040 & .233 & 288 \\
\bottomrule
\end{tabular}%
}
\end{table*}

\begin{table}[htbp]
\centering
\caption{Subjective-outcome summaries after Task~1}
\label{tab:subjective-summary}
\small
\begin{tabular}{@{}L{3.2cm}C{1.2cm}C{1.2cm}L{2.8cm}C{1.1cm}@{}}
\toprule
Outcome & Control $M$ & Scaffold $M$ & $b$ [95\% CI] & $P$ \\
\midrule
AI helpfulness & 4.77 & 4.40 & $-0.342$ [$-0.510$, $-0.174$] & $<.001$ \\
Perceived control & 4.09 & 3.78 & $-0.251$ [$-0.486$, $-0.017$] & .036 \\
\bottomrule
\end{tabular}
\end{table}

\end{document}